\documentclass[logo,msc,ai]{infthesis}         
\usepackage{msccheck}

\usepackage{microtype} 
\usepackage{amsmath}
\usepackage{tabularx}
\usepackage{float}
\usepackage{tikz}
\usepackage{pgfplots}
\usepackage{enumitem}
\usepackage{filecontents}
\usepackage{url}
\begin{document}
\begin{preliminary}

\title{Empirical Analysis of the Effect of Context in the Task of Automated Essay Scoring in Transformer-Based Models}

\author{Abhirup Chakravarty}

\submityear{2023}

\abstract{
Automated Essay Scoring (AES) has emerged to prominence in response to the growing demand for educational automation. Providing an objective and cost-effective solution, AES standardises the assessment of extended responses. Although substantial research has been conducted in this domain, recent investigations reveal that alternative deep-learning architectures outperform transformer-based models. Despite the successful dominance in the performance of the transformer architectures across various other tasks, this discrepancy has prompted a need to enrich transformer-based AES models through contextual enrichment.

This study delves into diverse contextual factors using the ASAP-AES dataset, analysing their impact on transformer-based model performance. Our most effective model, augmented with multiple contextual dimensions, achieves a mean Quadratic Weighted Kappa score of 0.823 across the entire essay dataset and 0.8697 when trained on individual essay sets. Evidently surpassing prior transformer-based models, this augmented approach only underperforms relative to the state-of-the-art deep learning model trained essay-set-wise by an average of 3.83\% while exhibiting superior performance in three of the eight sets.

Importantly, this enhancement is orthogonal to architecture-based advancements and seamlessly adaptable to any AES model. Consequently, this contextual augmentation methodology presents a versatile technique for refining AES capabilities, contributing to automated grading and evaluation evolution in educational settings.

}

\maketitle

\begin{acknowledgements}
I would like to express my absolute gratitude to my supervisor, Alexandra Birch, whose invaluable guidance, insightful feedback, and timely, yet unwavering encouragement made this research a success. 

I am indebted to the University of Edinburgh for providing the necessary resources, library facilities, and academic environment that fostered a conducive atmosphere for research and learning.\\
My friends and family deserve special mention for their unending support, encouragement and understanding during this demanding journey. Your unwavering belief in me provided the motivation to persist through challenges.\\
Lastly, I would like to acknowledge all the authors, researchers, and scholars whose works have been cited in this dissertation. Their contributions formed the foundation upon which this research was built.\\
I would like to acknowledge the contributions and support of numerous individuals and institutions without whom the completion of this dissertation would not have been possible.\\
Thank you.

\end{acknowledgements}

\tableofcontents
\end{preliminary}

\chapter{Introduction}
\label{intro}
Automated Essay Scoring (AES) has emerged as a solution in the wake of the digital revolution, where technology reshapes education. This digital paradigm shift has paved the way for various innovations in educational assessment. It seeks to automate the traditionally time-consuming and subjective process of essay grading. It is driven by the desire to expedite and standardise grading procedures, enabling educators to allocate more time to focused teaching and personalised learning experiences. Thus, as the demand for scalable assessment solutions rises, AES becomes a promising solution to this challenge. 

However, it is important to note that the effectiveness of an AES system hinges on the underlying algorithms and models. There has been significant research using different model architectures in the AES task, and although transformer-based models have shown remarkable success in various natural language processing tasks \cite{min2021recent}, their performance in AES tasks has revealed certain limitations. This was demonstrated by Tashu \textit{et al.} \cite{tashu2022deep}, in contrast to the limitations discussed by Wang \textit{et al.} \cite{wang_2022_on}. By using a Convolution Neural Network (CNN) and Recurrent Neural Network (RNN) based architecture, Tashu's DeLAES model significantly outperformed the transformer-based state-of-the-art model of Yang \textit{et al.} \cite{yang-etal-2020-enhancing}. It is important to note that none of these transformer-based models implemented proper contextual support to leverage the data fully. This inadequacy sparks the need to explore context augmentation techniques that could potentially enhance the capabilities of transformer-based AES models. 

Contextual cues have been found to impact human understanding of language significantly \cite{FERSTL202137}, and there is a growing recognition of their potential in improving AI-driven language tasks\cite{Hung2014}. Our research aims to harness this potential by investigating the integration of context in the context of AES. At the core of this research lies a fundamental question: 
\begin{quote}
    How can we enhance transformer-based models for AES by infusing them with contextual information? 
\end{quote}
Addressing this question necessitates a deep exploration of the role of context in the AES domain, as well as some other open-ended questions:
\begin{itemize}
    \item What is the point of an essay?
    \item What makes an essay better? 
    \item What does it mean to have a relevant context? 
    \item What kind of context may help in the downstream task?
\end{itemize}
We assert that an essay can be described as a textual composition crafted to persuade the reader of the author's perspective. Consequently, essays primarily strive to argue for a specific standpoint concerning the topic at hand. We substantiate this based on Mercier and Sperber's research \cite{mercier_sperber_2011}, which underscores the intrinsic argumentative nature of human reasoning, further reinforcing the inherently persuasive quality of essays. 

To restrict our scope of research, we limited our search to the following granularities: Relative Context, Prompt Context, Structural Context and Feature-based Context.
The essays are graded on a curve, implying that an individual essay would have a relative rank amongst all the other essays for a specific prompt. Margin Ranking, as discussed in Section \ref{sec:margin_ranking}, explores this in the form of relative context.
As discussed in Section \ref{subsec:asap}, the dataset has essays collected from eight different sets where every essay is written in response to their respective prompt. Understanding the prompt or the topic should be crucial to assess the essay adequately. This should give the model an insight into the essay's key elements, ideally better identifying unique patterns for good and bad aspects of a given prompt. Thus, a prompt can successfully act as a proper context.
Essays may follow different genres and writing styles, such as argumentative, narrative, or expository. However, we assume that an essay's primary objective is to convince the reader of some truth. To this end, if we provide the latent structural context of the essay to the model, it may understand the interdependencies of the sentences better. To address this, we want to introduce the structural contexts at the scale of Elementary Discourse Units (EDUs) \cite{rst-edu} and Argument Components (ACs) \cite{lawrence-reed-2019-argument}.
As discussed in Section \ref{asap-bg}, the length is correlated with the scores in the dataset. In that case, we argue that we should also use essays' string length and word count as quality indicators for the available data. Other contexts that may be useful could be the counts of EDUs and ACs identified in an essay to further act as an indicator of quality for the model.

Our objective is twofold: to systematically investigate the impact of context on the performance of transformer-based models in AES and to propose methods for enhancing these models through context augmentation. Our hypothesis posits that incorporating context will lead to improved AES performance. Context-rich transformer-based models will better capture the nuances of essays, enabling them to accurately evaluate written content across a range of prompts and argumentative structures. Thus, this dissertation is structured to explore the context-augmented AES landscape.  In essence, this research endeavour navigates the evolving terrain of automated essay scoring, seeking to harness the power of context to propel the capabilities of transformer-based models. Through meticulous investigation and empirical validation, we aspired to contribute to AI in education, advancing assessment methodologies for the digital era.

Combining the contexts from these levels for a composite context augmentation led to us achieving a mean Quadratic Weighted Kappa (QWK) score of 0.8213 when trained on all essay sets and 0.8697 for prompt/essay-set specific training. In line with our primary goal, we enhanced performance in transformer-based models with context augmentation and outperformed all the models available in the literature. DeLAES still beats our prompt-specific implementation with composite context augmentation by approximately 3.83\% but fails to do so in all essay sets. Another thing to note is that although we do not outperform DeLAES, our progress is non-disjoint from DeLAES's. Their progress is primarily attributed to the architectural soundness, whereas ours is done in input enrichment with context augmentation; this can be stacked over DeLAES' implementation.
The subsequent sections will delve into the existing literature in the Background section, briefly reviewing existing literature in AES, especially in the context of transformer-based models, the challenges thus faced, and more relevant information. Following this, the Methodology section will detail the proposed approaches to incorporating context and enhancing transformer-based models for AES tasks. The Experiments section will present the empirical investigations conducted to validate the hypothesis, while the Results section will analyse the outcomes of these experiments. The Conclusions section will synthesise the findings and offer insights into the implications of context augmentation in AES.

\chapter{Literature Review}
\label{sec:bg}
Before we expand on the research, we must establish a background to the backdrop of the preexisting literature in Automated Essay Scoring (AES) and other correlated information about our work. In this chapter, we would like to break it down into simple sections addressing the precursor knowledge about the field of AES, the current approaches, and related material that we will use to motivate our approach.

\section{Background}
As discussed before, AES has emerged as a prominent field, with specific critical contributions since the 1950s shaping its development. Ellis Batten Page's work has been widely acknowledged as a fundamental pillar in the advancement of AES \cite{page2003project}. As early as 1966, Page advocated for the feasibility of computer-based essay scoring \cite{page66}. In 1968, he published his influential work on Project Essay Grade (PEG), a specialised programme tailored for essay evaluation \cite{page68}. Subsequently, Page collaborated with esteemed institutions like the Educational Testing Service (ETS), engaging in continuous refinement and updates to PEG, which culminated in successful trials in 1994 \cite{page1994}.

Another noteworthy contribution to AES came from Peter Foltz and Thomas Landauer, who introduced the Intelligent Essay Assessor (IEA) in 1997 to score essays in undergraduate courses \cite{Foltz1999}. IEA has since evolved into a product offered by Pearson Educational Technologies and has gained extensive usage across commercial applications and state and national exams \cite{PearsonAssessmentsK12}.

Within this landscape, ETS played a pivotal role in advancing AES by developing "e-rater," an automated essay scoring program first employed in commercial settings in February 1999 \cite{Burstein2003TheES}. Jill Burstein, a prominent figure, took the lead in spearheading the team responsible for its development. ETS's Criterion Online Writing Evaluation Service leverages the e-rater engine to provide users with comprehensive scores and targeted feedback on their writing performance.

\section{Automated Student Assessment Prize}
\label{asap-bg}
The Hewlett Foundation organised the Automated Student Assessment Prize (ASAP) in collaboration with Kaggle in 2012 to evaluate AES systems' reliability. It predicted human raters' essay scores \cite{asap-aes}. The competition attracted 201 participants and featured a demonstration with nine AES vendors on a subset of the ASAP dataset. The ASAP-AES dataset remains the critical resource for AES research \cite{asapreview}, including our study.

However, the claim of AES equivalence to human raters has drawn scrutiny \cite{PERELMAN2014104}. Concerns include methodology issues like paragraph-based datasets, reliance on human graders for specific datasets, and use of the "resolved score" instead of the true score \cite{Hamner2012ContrastingSA}. Addressing these points is essential for robustly evaluating AES systems.

Jeon and Strube \cite{jeon-strube-2021-countering} noted a length-score correlation in the ASAP-AES dataset except for one prompt. Correlation coefficients were high (Pearson: 0.702, Spearman: 0.707, p-values $<$ 0.001). Perelman \cite{PERELMAN2014104} also raised this bias in 2014.

\section{Current Approaches}
To give a brief overlay of the current approaches, we will shortly discuss some recent papers that significantly impacted the scientific discourse in AES. In 2018, Ye Tai \textit{et al.} \cite{tay_2018_skipflow} introduced a novel approach, incorporating neural coherence features into an end-to-end model for text scoring. Their system outperformed existing methods on benchmark datasets, demonstrating the need for more accurate and robust approaches. In the same year, contemporary limitations were addressed by Cozma \textit{et al.} \cite{cozma-etal-2018-automated} by utilising string kernels to extract features from essays and word embeddings to capture semantic meaning. Their approach exhibited greater flexibility and adaptability in handling diverse text types and capturing nuanced aspects of language use.

Another interesting take in 2021 was that of Ma \textit{et al.}, who modelled essays' hierarchical structure \cite{Ma2021}. They proposed a novel approach that utilises a hierarchical graph structure based on Graph Convolutional Networks (GCN) to encode the document structure of essays. By leveraging graph aggregation, their method captures structured coherence and discourse information. Experimental evaluations on the ASAP dataset demonstrate the effectiveness of their approach in enhancing AES. This implied that using GCNs could benefit the AES task if the latent hierarchical structure in essays can be exploited well.

A novel approach to AES came with the implementation of the concept of Multi-Task Learning (MTL) as introduced by Kumar \textit{et al.} \cite{kumar-etal-2022-many} for AES systems, as opposed to Single-Task Learning (STL). Their study demonstrated the superiority of MTL systems utilising Long Short-Term Memory (LSTM) and Bi-Directional LSTMs (Bi-LSTMs), outperforming STL systems. Wang \textit{et al.} in 2022 \cite{wang_2022_on} employed BERT in a joint learning framework, capturing multi-scale representations of essays to enhance performance. They utilised various loss functions and transfer learning techniques, including out-of-domain data, to further improve the system's capabilities. Their work also introduced a custom loss function called Margin Ranking, discussed later in Section \ref{sec:margin_ranking}. This helps balance out the model during training as it provides contextual ordering of the essays present in the batch based on their score. Table \ref{table:comparing_results} shows their best-performing model's scores on the dataset.

The limited, somewhat noisy data was one of the primary reasons, as pointed out by Wang \textit{et al.} in their work which results in the sub-par performance of transformer-based models. Jong \textit{et al.} \cite{jong2022improving} addressed this issue of limited diversity in training data by proposing augmentation techniques such as back translations and adjusted scores for AES. Yang \textit{et al.} \cite{yang-etal-2020-enhancing} proposed a method for fine-tuning pre-trained language models using multiple losses. Their approach combined mean square error and batch-wise ListNet losses to enhance score calculation. They called it a combination of regression and re-ranking, and their model was thus called $R^2$-BERT as noted in Table \ref{table:comparing_results}. Theirs was also the model that held the state-of-the-art performance before Tashu \textit{et al.}'s approach was implemented.

More recently, Tashu \textit{et al.} \cite{tashu2022deep} introduced an innovative approach by using Multi-Channel Convolutional Neural Networks (CNNs) together with Recurrent Neural Networks (RNNs). Their architecture was named DeLAES and was trained on prompt-specific data, i.e., trained separately on each essay set. This helped them model the n-gram attributes (2-grams, 3-grams and 4-grams) from the word embeddings. These were then consolidated into comprehensive essay-level vectors utilising the process of max-pooling. Additionally, the architecture incorporates the utilisation of Bidirectional Gated Recurrent Units (BGRUs), enabling access to antecedent and subsequent contextual representations. Their work established the architecture's significance in AES as the new state-of-the-art. The performance of their model is also noted in Table \ref{table:comparing_results}.
\section{Latent Structure}
This section will explore potential ways to leverage the latent structure inherent in text-based inputs. An Elementary Discourse Unit (EDU), which is a sub-sentence phrase unit, was first proposed from Rhetorical Structure Theory (RST) authored by Mann and Thompson \cite{rst-edu}. It offers a holistic context level in that it acts as a complete unit of meaning and thus could help a model identify a complete thought or idea. 
A more succinct context for our use case would be that of an Argument Component (AC). The analysis of argumentation within discourse holds a critical role in argumentation theory, and central to this analysis is the concept of ACs, which provides insight into the structural composition of arguments. Van Eemeren and Grootendorst's influential work \cite{pdafr} set a significant milestone in understanding argumentation analysis. While this seminal piece of literature does not explicitly introduce the term "Argument Component," it provides a structured framework for identifying and assessing the discrete foundational elements that constitute arguments.

Another important note is that, in developing a model to detect ACs, it's crucial to consider a concise study by Lawrence \textit{et al.} \cite{lawrence-reed-2019-argument}. Their research emphasises the correlation between the Rhetorical Structure Theory (RST) discourse structure and argumentative constructs. They show that the discourse relations in RST often align with argumentative relations. This sheds light on the potential to use EDU-level context to enhance a model's ability to understand the concept of ACs better. 

These insights could help our model learn the text from different levels of granularity, allowing it better to grasp the connection between argumentative relations in text and enhance its ability to analyse the latent structural complexities of an essay.

\chapter{Data}
The data required for this project involved three major datasets. ASAP-AES dataset \cite{asap-aes} was used for the Automated Essay Scoring (AES) task, the RST Discourse Treebank dataset \cite{carlson2002rst} was used for Elementary Discourse Unit (EDU) span labelling and the Argument Annotated Essays (V2) \cite{aae2} was used for Argument Component (AC) span labelling.

\section{ASAP-AES}
\label{subsec:asap}
The ASAP-AES dataset\cite{asap-aes}, developed by the Hewlett Foundation, is a pivotal resource in AES research. With diverse essays paired with human-scored responses, it facilitates the evaluation of AES algorithms. It comprises essays belonging to separate sets. This sectioning was done because each set of essays was written in response to a particular prompt or question specific to that essay set. 

A sample point from the data looks like this (only columns with values included):

\begin{table}[hbtp!]
    \centering
    \resizebox{400pt}{!}{
    \begin{tabular}{|c|c|p{6cm}|c|c|c|c|c|c|c|}
        \hline
        essay\_id & essay\_set & essay & rater1\_domain1 & rater2\_domain1 & rater3\_domain1 & domain1\_score \\
        \hline
        1 & 1 & Dear local newspaper, I think effects computers... & 4 & 4 & 0 & 8\\
        \hline
    \end{tabular}
    }
    \caption{Sample Essay Data from ASAP-AES}
    \label{tab:essay_data}
\end{table}

The dataset encompasses 12,876 scored essays distributed among eight essay sets, each characterised by distinct prompts or, loosely, questions. They are derived from educational institutions and exhibit diverse scoring ranges across different sets. Please refer to Section \ref{app:prompts} for a comprehensive list of these prompts.

Trained human raters scored essays based on rubrics, ensuring reliability. Essays' lengths and writing quality vary, reflecting real-world scenarios. It aligns with our research as a comprehensive benchmark to assess our automated essay scoring model's performance. The details about the dataset can be glanced from Table \ref{table:asap-aes-ds}

The benchmark of this dataset was established such that a model that attains the higher Quadratic Weighted Kappa (QWK) score (discussed in Section \ref{sec:qwk}) is better.

\begin{table}[hbtp!]
\centering
\resizebox{\linewidth}{!}{
\begin{tabular}{|c|c|c|c|c|c|c|c|c|c|}
\hline
\textbf{Essay Set} & \textbf{Type of Essay} & \textbf{Grade Level} & \textbf{Training Set Size} & \textbf{Min Domain 1 Score} & \textbf{Max Domain 1 Score} & \textbf{Min Domain 2 Score} & \textbf{Max Domain 2 Score} & \textbf{Max Tokenised Length}\\
\hline\
1 & Persuasive/Narrative/Expository & 8 & 1783 & 2 & 12 & \_ & \_ & 1075\\
2 & Persuasive/Narrative/Expository & 10 & 1800 & 1 & 6 & 1 & 4 & 1321\\
3 & Source Dependent Responses & 10 & 1726 & 0 & 3 & \_ & \_ & 463\\
4 & Source Dependent Responses & 10 & 1772 & 0 & 3 & \_ & \_ & 462\\
5 & Source Dependent Responses & 8 & 1805 & 0 & 3 & \_ & \_ & 766\\
6  & Source Dependent Responses & 10 & 1800 & 0 & 3 & \_ & \_ & 548\\
7  & Persuasive/Narrative/Expository & 7 & 1569 & 0 & 3 & \_ & \_ & 978\\
8  & Persuasive/Narrative/Expository & 10 & 723 & 0 & 3 & \_ & \_ & 1489\\
\hline
\end{tabular}
}
\caption{ASAP-AES Dataset}
\label{table:asap-aes-ds}
\end{table}

While informative, the dataset might only cover some writing scenarios and could exhibit some rater variability. Also, the publicly available dataset only has a scored training set, but the validation and test set are not scored.

\section{RST Discourse Treebank}
The RST Discourse Treebank (RST-DT) \cite{carlson2002rst} is a pivotal resource for discourse analysis, offering annotated text segments with discourse structure. Developed by Lynn Carlson, Daniel Marcu, and Mary Ellen Okurowski based on the RST framework of Mann and Thompson, it aids in studying text coherence and discourse relations. There is a total of 385 Wall Street Journal articles taken from the Penn Treebank divided into 347 documents in the training set and 38 in the test set. The input data consists of sentences split by a newline character, and the true output has the EDU text split by newline characters. A sample of this is as follows:\\
\textbf{Input:}
\begin{quote}
    Defense intellectuals have complained for years that the Pentagon cannot determine priorities because it has no strategy.
\end{quote}
\textbf{Output:}
\begin{quote}
    Defense intellectuals have complained for years\\
    that the Pentagon cannot determine priorities\\
    because it has no strategy.
\end{quote}

Based on the original work of Wang \textit{et al.} \cite{wang-etal-2018-toward}, we randomly sampled 34 articles (roughly 10\%) from the train set. Precision, Recall, and primarily F1-scores were used, as discussed in Section \ref{evalmetrics}. Our primary interests in this dataset were the raw text and the segmented EDUs for each essay individually.

\section{Argument Annotated Essays (version 2)}  

The Argument Annotated Essays (Version 2) dataset \cite{aae2} is a significant resource for studying argumentation structures in essays. It offers annotated essays with labelled argument components, fostering research in argument analysis.

Christian Stab and Iryna Gurevych curated the dataset, including diverse essays from various domains, each annotated with argumentative components. It facilitates the exploration of argumentation patterns and relations.

This consists of raw text and argumentative labels like the span of each AC, whether it is a Claim, Major Claim, or Premise, their interlinkage and the type of linkage being in Support, Attack, etc. There are a total of 322 train and 80 test essays. For this dataset, we also split 10\% of the train set (32 essays) randomly to curate our validation set. Our primary interests in this dataset were the unlabelled raw text and the spans of the ACs, which we extracted for training. For pre-processing, our data was similarly formatted to the RST-DT dataset.

\chapter{Methodology}
\label{sec:metho}

To discuss the methodology implemented in this paper, we need to reiterate the primary research question \textemdash What can be done to provide transformers with relevant context such that it performs well in the downstream task of Automated Essay Scoring? As discussed in Chapter \ref{intro}, this raises several questions. 
Based on the work of Mercier and Sperber's \cite{mercier_sperber_2011}, we assume that an essay is essentially a textual piece arguing in favour of some standpoint and is trying to convince the reader of the same.
We also reiterate here that we restrict our scope to the following four types of context as discussed in Section \ref{intro}:
\begin{enumerate}
    \item Relative Context: This is used to inform the model of the relative ordering of quality of the essays in terms of scores obtained.
    
    \item Prompt Context: The prompts to which the essays were written in response should give the model an insight into the essay's inherent quality.

    \item Structural Context: The structural context looks to explore the latent structural patterns and qualities present in the essays in the two granularities: 
        \begin{enumerate}
            \item Elementary Discourse Units (EDUs)
            \item Argument Components (ACs)
        \end{enumerate}

    \item Feature-based Context: With this, we intend to exploit the various directly or indirectly extractable features from essays to inform the model further.
\end{enumerate}

Now that we have defined the scope of our research, we need to talk about the standard architecture used for this work. All our experiments would be run on a transformer-based model with a common classifier head architecture used to train on all eight prompts of the ASAP-AES dataset together and get both individual and mean scores for the prompts. The goal is to experimentally find out how different contexts when added to the input, can help the performance of our model. For this, we would be sequentially going through the four different types of context we want to explore, as discussed above and observe their relative performance. 

\begin{figure}[hbtp!]
    \centering
    \hspace*{-2cm}\resizebox{300pt}{!}{
        \includegraphics{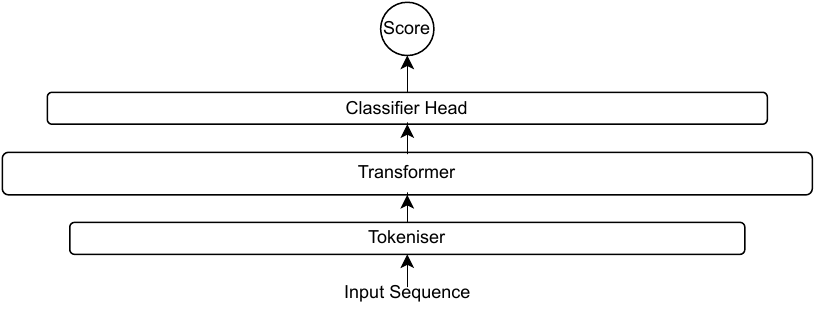}
    }
    \hspace*{-0.5cm}\caption{AES Model Architecture}
    \label{fig:aesmodel}
\end{figure}

Our initial candidate for the transformer architecture was RoBERTa \cite{liu2019roberta}. While BERT was the standard architecture for all of the discussed transformer-based models used for AES in Section \ref{sec:bg}, RoBERTa performs better than BERT across several benchmarks, as the source paper claimed. It was trained using a more extensive and diverse dataset with longer training times. The enhanced pre-training of RoBERTa helps it capture nuanced contextual relationships in language, making it well-suited for tasks that require a deep understanding of meaning and context \textemdash precisely what is needed for our use-case. It also incorporates techniques like dynamic masking and larger batch sizes during pre-training, leading to better regularisation and less overfitting. The decision to use a standard classifier head was made such that we could easily compare the performance of the baseline model with those of the context-augmented data, which involved changes to the architecture parameters that were minimal (only in the case of feature-based contexts) to none (in every other scenario). The architecture is demonstrated graphically using Figure \ref{fig:aesmodel}.

To clarify, the goal of this group of experiments is to find the effect of the augmentation of each of the types of context discussed below into the transformer model and see how it affects the performance. We also aim to combine these contexts to align the model with our goal of getting a high-performing AES system that can prove that transformer-based models can be considered worthy and capable of the AES task.

\section{Common Classifier Head}
Before proceeding further, we must adequately establish the architectural design choices made regarding the model. Since the transformer architectures we will be using are commonly available and widely discussed in the existing literature, in this section, we aim to describe the common classifier head we use across all our experiments. The function of the classifier head is to take as input the transformed embeddings of the input essay from the transformer architecture and output the predicted score for the input essay. The output of the transformer architecture is two-dimensional, with the first dimension representing the input sequence and the second dimension being the embedding dimension. All BERT-based models have an input dimension of 768, including all the models discussed in this work.

\begin{figure}[htbp!]
    \centering
    \hspace*{-2cm}\resizebox{230pt}{!}{
    \includegraphics[width=0.90\textwidth,height=500pt]{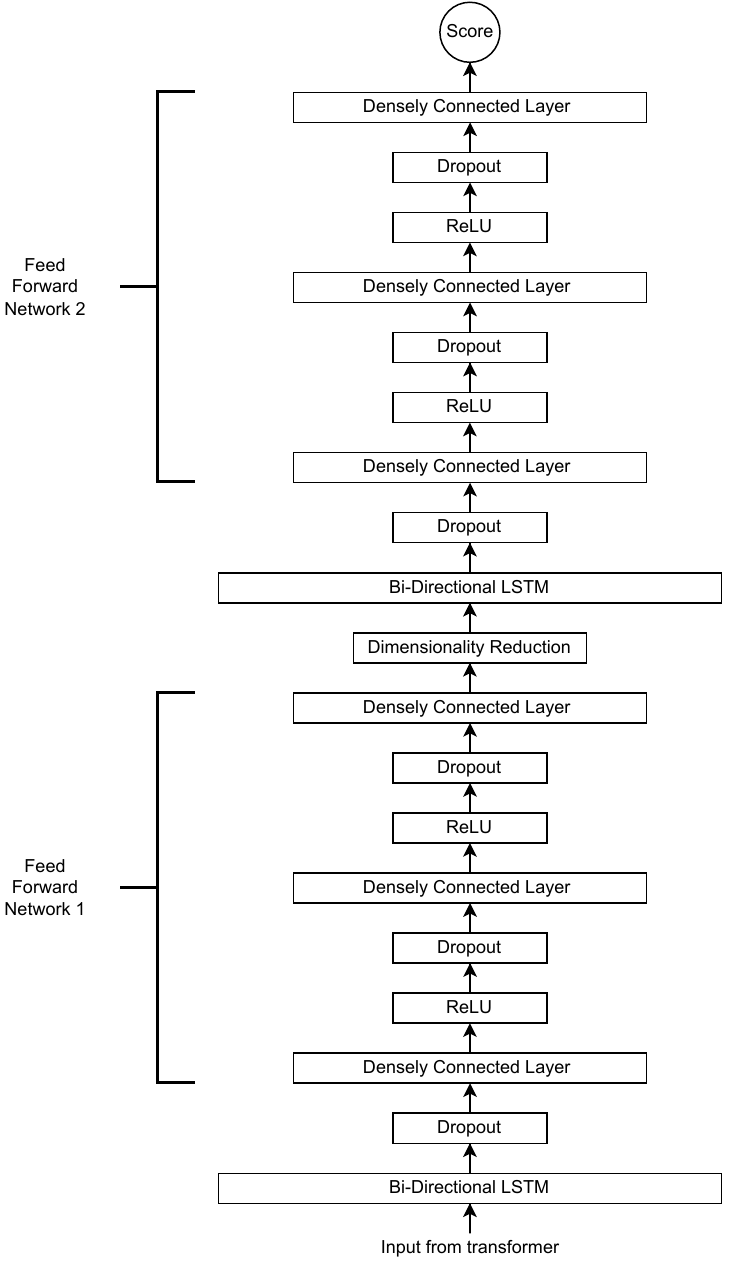}
    }
    \hspace*{-0.5cm}\caption{AES Model Architecture}
    \label{fig:classifier_architecture}
\end{figure}

We employ a combination of a Bidirectional LSTM (Bi-LSTM) and a composite Feed Forward Network (FFN) as the critical components of our classifier head. Specifically, we utilise this combination to process each input dimension separately and iteratively. The Bi-LSTM layers offer the best viable option to decode the sequential information generated from the transformers. The FFNs share a similar architecture, consisting of three densely connected layers. Dropout layers and ReLU activation provide regularisation and stability, respectively. The output size of the final layer of each FFN is set to 1, aligning with our prediction objective. This architecture is further clarified in Figure \ref{fig:classifier_architecture}.

These FFNs are equipped to handle each output dimension after the transformer architecture transformation. This distinctive design choice ensures that the FFNs adapt flexibly to the varying sizes of the two dimensions. Notably, after the first pair of FFN and Bi-LSTM, the output undergoes a dimension reduction step. This step is crucial as it prepares the output for processing by the second Bi-LSTM and FFN pair, which operates over the input sequence dimension. 

Typically the classifier head of a transformer-based model used for the AES consists of a simple dense layer or conjunction of dense layers and LSTMs (or Bi-LSTMs) \cite{wang_2022_on} \cite{yang-etal-2020-enhancing} \cite{liang2021rdrop}, so we decided that this design would represent performance as a standard design choice and the graded decrease in size in the FFNs would provide the model with a bit of stability.

To counteract the high correlation as found out by Jeon and Strube \cite{jeon-strube-2021-countering}, we use prompt-specific normalisation discussed in detail in Section \ref{scaler}. 

\section{Baseline}
The next question we must ask is \textemdash \hspace{1pt} how would the baseline perform without any added context? To answer this, we look into establishing a baseline with a transformer model. The architecture used is RoBERTa with the common classifier head. The results of this experiment can be seen in Table \ref{table:baseline-kappa}. In contrast to our design, most of the literature that uses transformer-based models have used BERT \cite{wang_2022_on} \cite{yang-etal-2020-enhancing} \cite{liang2021rdrop}, and some used Longformer\cite{beltagy2020longformer} as well \cite{wang_2022_on}. The classifier head for most of these implementations consisted of a simple, densely connected linear layer or a combination of LSTMs and dense layers. Although we use a standard architecture, our baseline's primary purpose is not to compare model implementations from literature but to provide a baseline score for later model performances. A point to note here is that since we use RoBERTa instead of BERT, we expect our baseline to outperform those in the preexisting literature.

\section{Input Length}
After establishing the baseline, we look at the next point of our interest \textemdash the maximum input length of the transformer model. As most previous implementations were based on BERT, the input sequence was truncated at the maximum sequence length of 512 tokens. Here, the question lies as such \textemdash how much does the truncation of the input affect the performance of the models? To understand this, since RoBERTa does not allow more than 512 tokens in the input sequence, we run an experiment establishing DeBERTaV3 with 512 tokens as the baseline for this. Since DeBERTaV3 uses relative position embeddings, training on the entire length of the input essays is possible. In the following experiment, we train on input data that pads to the maximum tokenised input length without any truncation. The results are further discussed in Section \ref{sec:results}, but it suffices to say that the non-truncated model outperforms the baseline DeBERTaV3 with truncation at the maximum length of 512 tokens. 

This made us reconsider our approach and choose Longformer \cite{beltagy2020longformer} for our transformer architecture. Longformer is based on RoBERTa, and we must note that RoBERTa does outperform DeBERTaV3. The modifications made to the Longformer architecture allow it to handle longer documents much better than RoBERTa. Thus, another baseline was established for Longformer using only raw data input. Table \ref{table:baseline-kappa} shows that the baseline Longformer model outperformed all others.

\section{Relative context: Margin Ranking}
\label{sec:margin_ranking}
Now that we have established that untruncated input would be ideal for transformer-based models, it is time for us to start looking at context augmentation, starting with relative context. To allow the model to understand the relative ordering of essays, we introduce Margin Ranking (MR) loss. MR loss was proposed by Wang \textit{et al.} \cite{wang_2022_on} to help the model understand the correct order of essays in a batch. The order of essays is crucial for scoring, as allowing the model to capture this aspect encourages the model to penalise incorrect ordering. For each batch of essays, we look at every possible pair of essays and calculate it as follows:

\[
MR(y, \hat{y}) = \frac{1}{\hat{N}} \sum_{i,j} \max \left(0, -r_{i,j} (y_i - y_j) + b \right)
\]

Where:
\begin{align*}
r_{i,j} &= \begin{cases}
1 & \text{if } \hat{y}_i > \hat{y}_j \\
-1 & \text{if } \hat{y}_i < \hat{y}_j \\
-\text{sgn}(y_i - y_j) & \text{if } \hat{y}_i = \hat{y}_j
\end{cases} \\
y_i & \hspace{1pt} \text{is the predicted score for the } i\text{ th essay}, \\
\hat{y}_i & \hspace{1pt} \text{is the actual score for the } i\text{ th essay}, \\
\hat{N} & \hspace{1pt} \text{is the number of essay pairs,} \\
b & \hspace{1pt} \text{is a hyperparameter set to 0 in our experiments.}
\end{align*}

In simpler terms, for each pair of essays (i, j), if the predicted score \(\hat{y}_i\) is higher than \(\hat{y}_j\), the actual score \(y_i\) should also be higher than \(y_j\); otherwise, we adjust the loss based on the difference between the scores. If the predicted scores are the same, the loss is based on the absolute difference between the actual scores. Based on the performance in Table \ref{table:added_context_kappa}, it was decided that MR loss should be used as an added context for the rest of the experiments.

\section{Prompt as context}
The inclusion of a prompt as an added context serves a dual purpose. Including the context, the model benefits from understanding the extent of this variance. The interaction between the essay and the prompt holds significance in assessing the essay's quality. The transformer model, when equipped with the ability to discern the nuanced relationship between the essay content and the prompt, gains the capability to recognise how the essay should ideally respond to and align with the given prompt. This comprehension empowers the model to evaluate better the essay's coherence, relevance, and overall merit.

The prompt is incorporated into the essay's textual representation to facilitate this interaction with special tokens that mark the demarcation between the original and appended content. Thus, this adjunction of the prompt with the essay content allows the model to effectively capture the interplay between them, ultimately enhancing its capacity to evaluate essays within the specific context of the given prompt. The augmentation of the prompt with the essay can be illustrated as follows:
\begin{quote}
    Sample Essay:\\
    Dear local Newspaper @CAPS1 a take all your computer and given to the people around the world for the can stay in their houses chating with their family and friend. Computers help people around the world to connect with other people computer help kids do their homework and look up staff that happen around the world.\\\\
    Added Prompt Context:\\
    $[PROMPT]$ We all understand the benefits of laughter. For example, someone once said, “Laughter is the shortest distance between two people.” Many other people believe that laughter is an important part of any relationship. Tell a true story in which laughter was one element or part. $[ESSAY]$ Dear local Newspaper @CAPS1 a take all your computer and given to the people around the world for the can stay in their houses chating with their family and friend. Computers help people around the world to connect with other people computer help kids do their homework and look up staff that happen around the world.
\end{quote}

\section{Elementary Discourse Unit as context}
\label{edu}
After exploring the relative ordering context, the next order of business is to start delving into the structural contexts. We will begin with the Elementary Discourse Unit (EDU) level of granularity.
\begin{figure}[hbtp!]
    \centering
    \hspace*{-2cm}\resizebox{300pt}{!}{
        \includegraphics{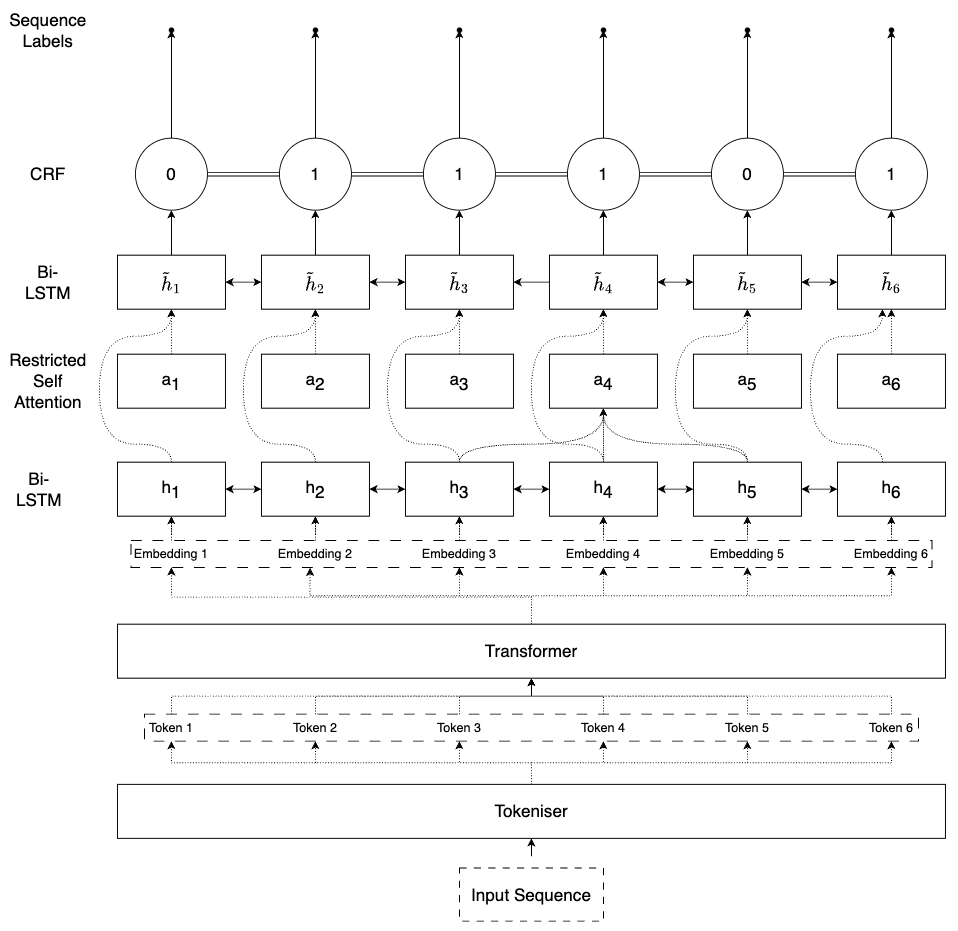}
    }
    \hspace*{-0.5cm}\caption{EDU Model Architecture}
    \label{fig:edumodel}
\end{figure}
As discussed, an EDU is the smallest unit of text with meaning. It is a concept from discourse analysis used to analyse the latent structure and organisation of text (discourse) on a sentence level. It helps understand the flow of information and the relationships between different text parts. They can range from single words to short phrases or clauses to a whole sentence. 

How it is planned for usage helps the transformer model understand individual parts of the text (units of meaning) and the information flow better. This is utilised by first segmenting all the essays with the EDU segmentation model we implement and then combining them with special tokens to signify the beginning of each EDU. This is then fed into the transformer to get embeddings. 

An illustration of the intended method can be made as such:
\begin{quote}
    Sample essay input:\\ 
    Dear local Newspaper @CAPS1 a take all your computer and given to the people around the world for the can stay in their houses chating with their family and friend. Computers help people around the world to connect with other people computer help kids do their homework and look up staff that happen around the world.\\\\
    EDU labelled input:\\ $[EDU]$ Dear local Newspaper @CAPS1 a take all your computer $[EDU]$ and given to the people around the world $[EDU]$ for the can stay in their houses $[EDU]$ chating with their family and friend. $[EDU]$ Computers help people around the world to connect with other people computer help kids do their homework $[EDU]$ and look up staff $[EDU]$ that happen around the world.
\end{quote}

\subsection{EDU Span Prediction}

Before using EDU-labelled data, we develop an EDU span classification model based on Wang \textit{et al.}\cite{wang-etal-2018-toward}. The reason we chose their work to base our architecture on is because of its high adaptation rate as a standard model to do the upstream task of EDU labelling \cite{ni-etal-2019-justifying} \cite{xu2020discourseaware} \cite{chen2020modeling} \cite{chakrabarty2020ampersand}\cite{saha-etal-2022-edu}. It is also open-source, allowing us to change the architecture as our needs are. The standard model also has quite an impressive F1-Score of 0.940. Figure \ref{fig:edumodel} describes the model architecture graphically. Our variation implements RoBERTa embeddings with the original BiLSTM-CRF network architecture.  RoBERTa was our choice of text embedding generator as it has a much better understanding of language than the original 300-D Glove embeddings. Also, there were no truncation concerns since it was used to get sentence-level embeddings and not document or essay-level. BiLSTMs address context loss in longer sequences. Long-range dependencies, crucial for EDU boundaries, are managed using Restricted Self Attention (RSA), discussed in Section \ref{rsa}. This was arguably the most essential contribution of Wang \textit{et al.} in the original paper. The Conditional Random Field used with the Bi-LSTM also significantly contributes to the performance superiority and is discussed in detail below.

\subsubsection{CRF}
Conditional Random Fields (CRFs) effectively model sequential dependencies, ensuring coherence and label consistency, and are well-suited for structured prediction, particularly sequence labelling. In linear-chain CRFs (as implemented), labels are assigned to sequence elements considering local features and inter-element relationships. The CRF model comprises three key components:

\begin{enumerate}
    \item Unary Potential (Node Potential):
The unary potential represents the compatibility of a label with the observed features at a single position in the sequence. It combines the influence of individual features on label assignments.
\[
\psi_u(y_i, x_i) = \exp \left( \sum_{k=1}^{K} \lambda_k f_k(y_i, x_i) \right)
\]

\item Pairwise Potential (Transition Potential):
The pairwise potential represents the compatibility between neighbouring labels in the sequence. It captures the dependencies between adjacent labels and enforces consistency in label assignments.
\[
\psi_p(y_{i-1}, y_i) = \exp \left( \sum_{l=1}^{L} \mu_l g_l(y_{i-1}, y_i) \right)
\]

\item Conditional Probability:
The conditional probability of a label sequence given the observed features can be computed using the product of unary and pairwise potentials, normalised by a partition function \(Z(X)\) to ensure valid probabilities.
\[
P(Y|X) = \frac{1}{Z(X)} \prod_{i=1}^{n} \psi_u(y_i, x_i) \prod_{i=2}^{n} \psi_p(y_{i-1}, y_i)
\]

\end{enumerate}

The unary potential captures label-element compatibility. Pairwise potential models neighbouring label relationships. By combining these, the CRF model integrates local-global info, predicting contextual labels coherently. CRFs handle context-dependent labels, modelling dependencies probabilistically. They provide structured predictions adhering to constraints. Training learns parameters \(\lambda_k\) and  \(\mu_l\) maximising data likelihood.
\subsubsection{Restricted Self Attention}
\label{rsa}
Standard self-attention captures global dependencies, but Wang \textit{et al.} showed EDU boundaries correlate mainly with nearby ones. Their solution, the Restricted Self-Attention (RSA) mechanism, compels the model to focus on nearby windows in stages, gathering pertinent information for problem-solving.

The formulation of restricted self-attention is somewhat like this:\\
Suppose that K is the hyperparameter window size. If $x_i$ is the current word and $x_j$ $\forall$ $j$ in the range $(i - K, i + K)$, then the first step is to compute the similarity between them as follows:
\begin{equation}
    s_{i,j} = w^{T}_{attn}[h_i,hj,h_i \cdot h_j]
\end{equation}
 where $h_m$ is the hidden representation of the word $x_m \forall m$ in sequence. To compute the attention vector $a_i$, we proceed as follows:
 \begin{equation}
     \alpha_{i,j} = \frac{e^s_{i,j}}{\sum^{K}_{k = - K} e^s_{i,i+k}}
 \end{equation}
 \begin{equation}
     a_{i,j} = \sum^{K}_{k = - K} \alpha_{i,i+k}h_{i,i+k}
 \end{equation}
This is used between two BiLSTM layers, and the attention vectors are concatenated with the previous BiLSTM layer to act as the input for the next layer as follows:
\begin{equation}
    \tilde{h}_t = BiLSTM(\tilde{h}_{t - 1}, [h_t, at])
\end{equation}
These outputs are then fed to the CRF for label predictions.

\section{Argument Component as context}
This section delves into context granularity via Argument Components (ACs) - distinct elements in discourse contributing to overall argument structure, including claims, premises, counterarguments, and evidence. ACs are fundamental in shaping the logical and persuasive structure of argumentative writing. We aim to identify and label ACs in the essay to enhance the transformer model's understanding of the writer's persuasive intent. This aligns with the essay's purpose of arguing for a standpoint. This exploration employs the same architecture (see Figure \ref{fig:aesmodel}) and labelling scheme as discussed in Section \ref{edu}. An illustrated example of the same is as follows:
\begin{quote}
    \textbf{Sample Essay:}
    Dear local Newspaper @CAPS1 a take all your computer and given to the people around the world for the can stay in their houses chating with their family and friend. Computers help people around the world to connect with other people computer help kids do their homework and look up staff that happen around the world.\\
    \textbf{AC-labelled Essay:}
    $[AC]$ Dear local Newspaper @CAPS1 a take all your computer and given to the people around the world for the can stay in their houses chating with their family and friend. [$AC]$ Computers help people around the world to connect with other people computer help kids do their homework and look up staff that happen around the world. 
\end{quote}

\subsection{AC Span Prediction}
As was with the EDUs as context, we also needed to implement a model for AC span classification. We decided to use the same architecture as our EDU span detection model. One thing to note is that we used EDU boundaries as signals to guide the model in learning the AC spans. We achieve this by adding special tokens to inform the model of the beginning and end of EDU spans within the text (similar to the illustrated example of an EDU-labelled essay).

The EDU span boundaries were used under advisement from the works of Peldszus and Stede \cite{Peldszus2015AnAC} \cite{peldszus-stede-2016-rhetorical}, as also Musi \textit{et al.} \cite{musi-etal-2018-multi} who demonstrated a correlation between discourse relations from Rhetorical Structure Theory (RST) and argumentative relations by translating the discourse structure mapping into argumentation structures. 

\section{Combined Contexts}
This section discusses how we can further combine the explored contexts to enhance the model's performance capabilities and bolster its reliability and correlation to human graders. After introducing and exploring  MR loss, we have already used it in all the following experiments. Now we consider combining the prompt context with the best performing structural context granularity \textemdash the context of Argument Component span. We also explore adding specific feature-based contexts that will be discussed in their respective subsection.

\subsection{Combining Prompts with ACs}
As discussed before, the AC spans and Prompts were combined with their special tokens to help the model understand how to process them separately and together. This conductive input is then provided to the transformer architecture. This should not only help the model figure out the relation between the prompt and the essay, and to a certain extent, the scoring ranges as well, but it will also demarcate the different claims and premises set up for the claims separately. An illustrated example can be seen as follows:

\begin{quote}
    \textbf{Sample Essay:} 
    Dear local Newspaper @CAPS1 a take all your computer and given to the people around the world for the can stay in their houses chating with their family and friend. Computers help people around the world to connect with other people computer help kids do their homework and look up staff that happen around the world.\\
    \textbf{Essay with combined context of Prompt and AC spans:} $[PROMPT]$ Write a letter to your local newspaper in which you state your opinion on the effects computers have on people. Persuade the readers to agree with you. $[AC]$ Dear local Newspaper @CAPS1 a take all your computer and given to the people around the world for the can stay in their houses chating with their family and friend. $[AC]$ Computers help people around the world to connect with other people computer help kids do their homework and look up staff that happen around the world. 
\end{quote}
\subsection{Argument Component with Prompts and Feature Context}

In this section, we discuss exploiting other feature-based contexts from the dataset.
For instance, we would like to refer to the discussion of the work of Jeon and Strube \cite{jeon-strube-2021-countering} from Section \ref{asap-bg}. They pointed out that the resolved score in the dataset is heavily correlated to the lengths of the input essays. 
\subsubsection{Analysis of Features}
Since we implemented prompt-specific scaling to ensure that the score distribution was fair and comparable amongst prompts, we wanted to do a statistical analysis of Pearson Correlation Coefficients between the variables directly available and some derived features, viz., AC Count (count of AC spans in an essay detected by the AC Span Labelling Model), EDU Count (count of EDU spans in an essay detected by the EDU Span Labelling Model), Essay Set, Length of Essay (essay's character length), Word Count, Score (resolved score from dataset), Scaled Score (prompt-specific normalised score).
\begin{figure}[hbtp!]
    \centering
    \resizebox{240pt}{!}{
        \includegraphics{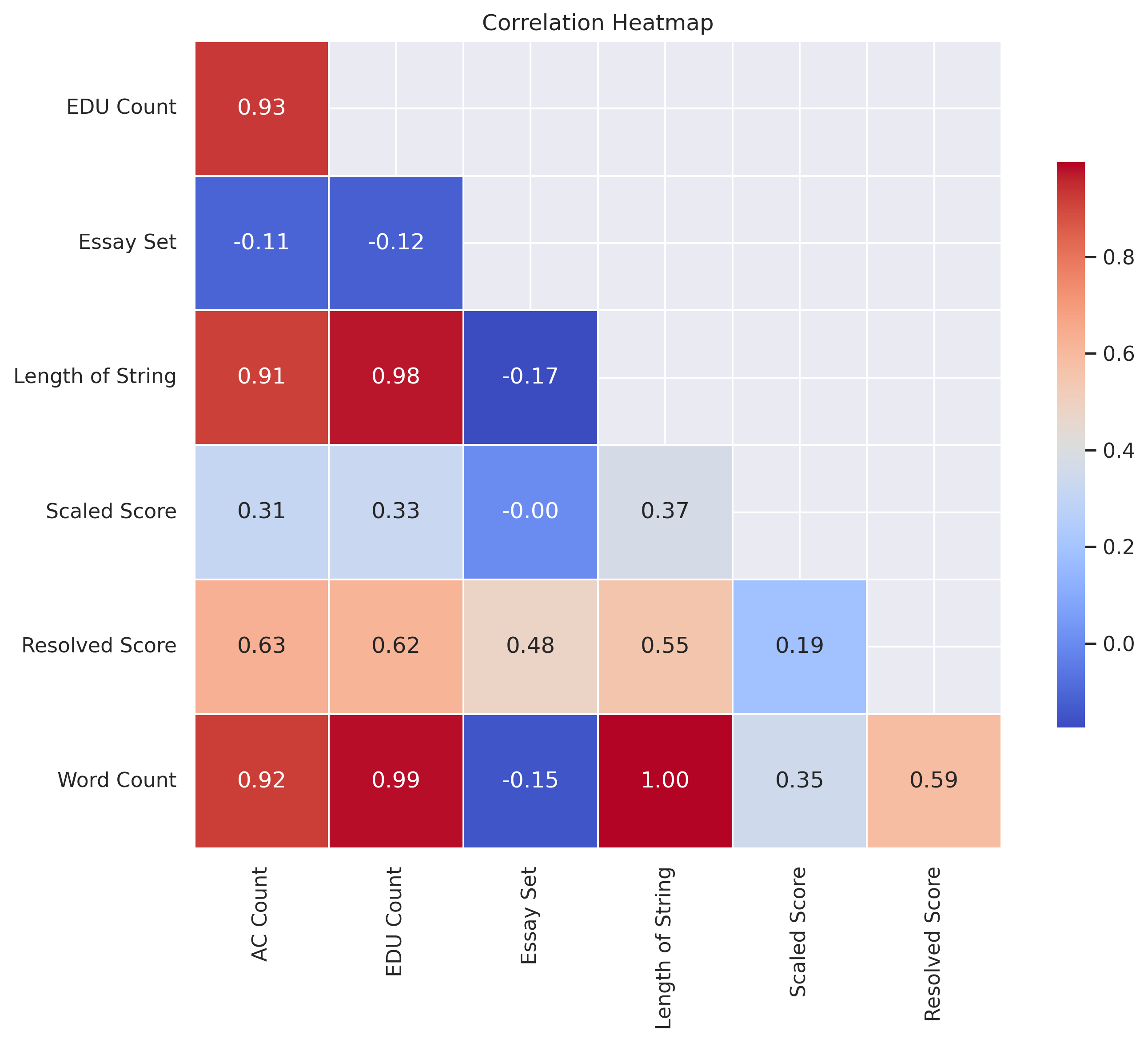}
    }
    \hspace*{1cm}\caption{Pearson Correlation Heatmap for Variables}
    \label{fig:corr}
\end{figure}
\label{sec:corr}\\
We saw that all the pairs had p-values $< 10^{-33}$, except for essay set and scaled score, which had a p-value of 1 (approximately).
The score or the resolved score from the dataset had a correlation with almost all the variables discussed, as was the case of the scaled score except between the essay set and itself. This is because the prompt/essay set specific normalisation of the score made it decorrelated to the essay set. The exact values are found in Figure \ref{fig:corr}. 

Since we observe a substantial correlation between the AC count, the EDU count, the Length of essays, and the Word Count of essays, it's essential to acknowledge that these variables are confounding factors. Each exhibits a correlation with the scaled score of over 0.3, indicating a positive moderate linear relationship. After careful consideration, we have chosen to incorporate all four of these variables in our analysis. While the confounding nature of these factors warrants vigilance, their individually strong correlations with the scaled score suggest that they collectively contribute meaningful information to our model's predictive capability.

So, we concatenated tensors of these three feature counts into the embedding tensors that act as the input to the second pair of LSTM and FNN. For example, for the sample essay:

\begin{quote}
    Dear local Newspaper @CAPS1 a take all your computer and given to the people around the world for the can stay in their houses chating with their family and friend. Computers help people around the world to connect with other people computer help kids do their homework and look up staff that happen around the world.
\end{quote}

We get a word count of 56, a length of the string of 317, an AC count of 2 and an EDU count of 7. As discussed above, we concatenate into one tensor [56, 317, 2,7] and attach this to the input tensor for the second LSTM-FNN pair.

\chapter{Experimental Design}
\label{sec:experi}
In this chapter, we will talk about the different design choices that we have made throughout our experiments and what motivated them. We will also discuss the models we want to compare our performances with. These choices range between evaluation metrics, loss functions, optimisers, hyperparameters, standardisation techniques and any other preprocessing we have done. We also briefly reintroduce the competing models discussed in Section \ref{sec:bg} and discuss why we deem them important.
\section{Evaluation Metrics}
In this section, we discuss the different metrics we employ to measure the performance of our models in the different training scenarios. 
The F1 Score was the metric of choice for the span labelling experiments. To reiterate, our goal with these experiments is to train a model to read text input and correctly classify the (positive) spans of EDUs and ACs, respectively. Since the classes are highly imbalanced, accuracy can be misleading because it might be high even if the model performs poorly in the minority class. F1 score considers both precision and recall, which are crucial for capturing performance across imbalanced classes. Also, our primary objective is to correctly identify and label the spans of interest. F1 score emphasises true positives (correctly identified spans) and true negatives (correctly not identified non-spans) while penalising false positives and false negatives. It is also a balanced view of both precision and recall. High precision indicates that the identified spans are mostly correct, and high recall suggests that most actual spans are being identified. The F1 score reflects the balance between these two aspects and is particularly useful when you need a compromise between precision and recall.

As for the ASAP-AES dataset, we used the Quadratic Weighted Kappa (QWK) score to benchmark our model's performance against others in the available literature. This is partially because this was the benchmarking tool advised by the ASAP contest on Kaggle and has been used as a benchmarking method for most if not all, papers on this dataset. However, elaborating further, the AES model aims to correlate scores predicted by the model with the human scores from the dataset.  QWK scores consider the model's ability to predict scores and reflect its consistency with human judgments. The essays in the dataset are scored on an ordinal scale, where the scores have a natural order but may not have a fixed interval between them. QWK handles ordinal data very well \cite{DELATORRE2018144} and assesses the agreement between different raters or systems based on these ordered categories.
Another thing to note is that the QWK is an extension of Cohen's Kappa metric, commonly used to enumerate inter-rater agreements. But as opposed to Cohen's Kappa metric, QWK assigns different weights to different levels of agreement/disagreement based on how far apart the scores are. This is particularly useful for our use case, where the distance between scores matters more than agreement or disagreement. It is also robust to chance agreements between raters. This is particularly useful as multiple human raters might score the same essay differently due to various factors. QWK adjusts for the possibility of chance agreement and provides a more accurate assessment of model performance. Like in many real-world scenarios, the ASAP dataset has non-linear scoring scales. The difference between a score of 3 and 4 might have a different impact than the difference between 7 and 8. QWK is better suited to capture these nuances.

The following subsections will briefly discuss the metrics more mathematically.

\label{evalmetrics}
\subsection{Precision}
Precision measures the proportion of accurately predicted positive instances out of all the cases predicted as positive by the model. It provides insights into the reliability of positive predictions. It is described mathematically in Section \ref{app:pre}.

\subsection{Recall}
Recall, also known as Sensitivity or True Positive Rate, quantifies the proportion of correctly predicted positive instances out of all actual positive samples. It assesses the model's ability to capture positive cases. It is defined mathematically in Section \ref{app:recall}.

\subsection{F1 Score}
The F1 score is a harmonic mean of precision and recall, providing a balanced evaluation of the model's performance. It proves valuable in scenarios with imbalanced class distributions where one metric needs more emphasis than the other. It is mathematically defined in Section \ref{app:f1}
As discussed before, it offers a comprehensive perspective on the classification model's effectiveness, showcasing its capacity to correctly identify positive instances, mitigate false positives, and strike a harmonious balance between precision and recall.

\subsection{Quadratic Weighted Kappa (QWK)}
\label{sec:qwk}
The QWK statistic quantifies the agreement by considering the difference between the observed and expected agreement, normalised by the maximum possible agreement. It assigns different weights to disagreements based on their magnitude, with a higher weight for more significant disagreements. The formula for calculating Quadratic Weighted Kappa is as follows:
\[
\mathrm{QWK} = 1 - \frac{\sum_{i,j}w_{ij}O_{ij}}{\sum_{i,j}w_{ij}E_{ij}}
\]
Where:\\
\resizebox{\linewidth}{!}{
\begin{minipage}{\linewidth}
\[
\begin{aligned}
    O_{ij} & : \text{Observed agreement between raters for each score category pair} \\
    &= \sum_i \sum_j w_{ij} \times \text{Instances with score } i \text{ by both automated system and human raters}
\end{aligned}
\]
\end{minipage}
}
\\
\resizebox{.85\linewidth}{!}{
\begin{minipage}{\linewidth}
\[
\begin{aligned}
    E_{ij} & : \text{Expected agreement under chance for each score category pair} \\
    &= \frac{\sum_i (\text{Instances with score } i \text{ by automated system}) \times (\text{Instances with score } j \text{ by human raters})}{\text{Total instances}}
\end{aligned}
\]
\end{minipage}
}\\
\resizebox{\linewidth}{!}{
\begin{minipage}{\linewidth}
\[
\begin{aligned}
    w_{ij} & : \text{Weighting matrix representing importance of each disagreement} \\
    &\rightarrow \text{Assigns varying weights to score category pairs based on significance}
\end{aligned}
\]
\end{minipage}
}\\

This formulation integrates the observed and expected agreements, accounting for the importance of disagreements through the weighting matrix. The \textit{QWK} offers a comprehensive measure of the correlation between the automated scoring system and human raters, providing insights into the agreement's nuances.

The weighted nature of QWK allows for a more nuanced evaluation of the model's scoring predictions, considering both the degree of disagreement and the specific score categories.

\section{Experimental Setup}
In this section, we will elaborate on the loss functions, optimisers, hyperparameters used and any other noteworthy details of the experiments that were run but more specific to the practical implementation than based on theory. For AES, the average over ten randomly sampled seeds for the best-performing model from the 10-fold cross-validation is presented in Section \ref{sec:results}. Since the Span Labelling tasks were treated as an upstream task, we saved the best performant model over ten seeds and reported the scores from the models we used to label the EDU and AC spans.

\subsection{EDU and AC Span Labelling}

The same parameters were used to train the model architecture used in EDU and AC labelling. The RST Discourse Treebank provides segmented EDU spans separately, whereas the AAE V2 dataset directly gives us the character-level spans of the ACs. These generated IO (Inside Outside) labels for the two tasks from the data. Negative Log Likelihood loss was used to train the models (for EDU and AC separately) along with L2 regularisation, and Adam \cite{kingma2017adam} was used for the optimiser. F1-score was used as the final metric to judge the efficacy of these models. The model architecture used is as described in Figure \ref{fig:edumodel}.

\subsubsection{IO Labelling}
IO labelling, or "Inside-Outside" labelling scheme, is a method of labelling scheme used for NER and Span detection tasks where if the text (token) is in the required span, it is labelled "Inside" or "I" or "Outside" or "O". It is similar to the tagging methods like "BIO" (Begin-Inside-Outside) or "BIOE" (Begin-Inside-Outside-End). Still, it is the simplest of all of these.  For our use cases, every token in an EDU (for the EDU span labelling training) or an AC (for the AC span labelling training) will be marked "I", and all the other tokens will be marked "O", and the model needs to predict this tag to get the correct prediction for that token.
\subsubsection{Negative Log Likelihood}
Negative Log Likelihood (NLL) measures the discrepancy between the predicted class probabilities and the actual class labels. It quantifies the model's ability to predict the next element in a sequence based on the previous ones. Minimising the negative log-likelihood during training is equivalent to making the model's predictions as close as possible to the observed data. In a sense, it measures how well the model fits the data distribution. To break it down, we need first to consider the likelihood. Likelihood estimates how well a given model explains observed data. It tells us how probable an observed data is under a particular parameterised condition or distribution. Since likelihood values can become significantly small in real-world applications, working with their logarithm values is more convenient. A natural log of the likelihood also converts its products to sums and simplifies the computational power required for the calculations.

NLL is used as an objective function for optimisation problems where minimising it is equivalent to maximising the likelihood, making it more probable for the model to predict the observed data \textemdash which in our case is the training labels. Thus, reducing the negative log-likelihood gives the model a higher probability of predicting the correct and desired labels for a given sequence of tokens. The mathematical formula for NLL is defined in Section \ref{app:nll}

\subsubsection{L2 Regularisation}
L2 regularisation, also known as Ridge regularisation, prevents overfitting in machine learning models by adding a penalty term to the loss function that discourages large values of the model's parameters. This is achieved by adding the squared L2 norm (Euclidean norm) of the model's parameters to the loss. A point to note is that L2 regularisation and Mean Squared Error (MSE) are distinct yet interrelated concepts. L2 regularisation is a technique aimed at mitigating overfitting by adding a penalty proportional to the squared L2 norm of the model's parameters to the loss function, effectively encouraging smaller parameter values. On the other hand, MSE serves as a loss function used for training and evaluating regression models, quantifying the average squared difference between predicted and actual values. While both contribute to improving model performance, L2 regularisation focuses on controlling parameter magnitudes to prevent overfitting, while MSE assesses prediction accuracy by directly measuring prediction discrepancies. The formula for L2 regularisation (Ridge regularisation) is described in Section 
\ref{app:l2}
\subsubsection{Adam}
\label{exdes:adam}
Adam \cite{kingma2017adam} (Adaptive Moment Estimation) is an optimisation algorithm. It combines the advantages of both Adagrad \cite{adagrad} (which adapts learning rates based on historical gradient information) and RMSProp \cite{tieleman2012lecture}  (which uses moving averages of squared gradients) to provide fast and effective optimisation.

It maintains moving averages of both the gradient and the squared gradient (also known as the second moment). It adapts the learning rate for each parameter by considering both the gradient and the past gradients' magnitudes. The mathematical formulae for Adam can be referenced from the Appendix in Section \ref{app:adam}.

\subsubsection{Hyperparameters}

\begin{table}[htbp!]
\centering
\begin{tabularx}{\textwidth}{|X|c|}
\hline
Hyperparameter & Value \\
\hline
Learning Rate & 0.001 \\
Weight Decay & 1e-4 \\
Dropout Keep Probability & 0.9 \\
Exponential Moving Average Decay & 0.9999 \\
Max Gradient Norm & 5.0 \\
Batch Size & 32 \\
Epochs & 50 \\
Window Size & 5 \\
\hline
\end{tabularx}
\caption{Hyperparameters for Span Labelling model}
\label{table:hyp_span}
\end{table}

The same hyperparameters were used to train the span labelling models. These hyperparameters used can be referred from Table \ref{table:hyp_span}.

\subsection{Automated Essay Scoring}
In the Automated Essay Scoring task, just Mean Squared Error loss was initially used, but later, the combined weighted loss of MSE and Margin Ranking (MR) was implemented. RMSProp\cite{tieleman2012lecture} was used as the optimising algorithm, and $scikit-learn$'s $StandardScaler$ was used to normalise the scores. The early stoppage technique was used to save computation time if the validation QWK score did not improve in 15 epochs, and validation QWK scores were used to keep the best-performing model.

\subsubsection{Score Scaling}
\label{scaler}
The score scaling (Z-score normalisation) scales the scores (which have a varied range in our dataset, based on different prompts) such that it has an overall mean of 0 and a standard deviation of 1. Thus, the distribution is converted into a standard normal distribution (zero mean and unit variance). A point to note here is that the Z-score normalisation was done in a prompt-specific manner such that the different essays could be scored on a similar type of distribution and reduce the noise in the data due to the varying score ranges along the dataset. This is in accordance with what was discussed in Section \ref{sec:corr}.

The scaling process begins by "centring" data in which the mean is subtracted from each data point. Thus, the data now has a mean of zero. This helps optimisation algorithms converge faster during training and improves the model's overall performance and stability. The centred data is divided by a standard deviation, ensuring the features have similar scales. This further prevents scores with large values from disproportionately affecting the training. This is mathematically shown in Section \ref{app:norm}

\subsubsection{MSE Loss}
The Mean Squared Error (MSE) loss function was employed to evaluate our trained deep neural networks' performance. MSE loss is a common metric used in regression tasks, including our study's predictive modelling objectives. The simplicity and intuitiveness of MSE make it a suitable choice for quantifying the discrepancy between predicted and actual values. MSE guides the training process towards achieving more accurate predictions by minimising the squared differences. Its clarity and alignment with our research goals make MSE an apt selection for assessing the effectiveness of our models in capturing the underlying relationships within the data. It is mathematically described in Section \ref{app:mse}

\subsubsection{Combined Loss}
Combined loss, in our case, is simply a weighted average of the MSE loss and MR loss (as discussed in the Methodology section), described mathematically as follows:

\[
\text{Combined Loss} = \alpha \cdot \text{MSE} + \beta \cdot \text{MR}
\]
where \( \alpha \) and \( \beta \) are weighted mean coefficients, thus  \( \alpha \), \( \beta \) $\in$ [0,1], and \( \alpha = 1 - \beta \). MSE represents the Mean Squared Error, and MR represents Margin Ranking loss.
\subsubsection{RMSProp}
\label{RMS Prop}
RMSProp \cite{tieleman2012lecture} was primarily chosen for the AES model for its stable and reliable convergence behaviour. Its gradual adjustments to learning rates make it particularly suited for non-convex optimisation problems, where finding the best solution is challenging. Unlike algorithms like Adam, which can lead to rapid changes in learning rates, RMSProp's approach avoids potential instability during training, ensuring a smoother learning process.

Another reason for opting for RMSProp is its controlled learning rate adaptation. This suits our goal of maintaining a consistent and slower adjustment of learning rates, which can be crucial for specific scenarios. The characteristics of our dataset were also taken into account. It is better equipped to handle noisy or sparse data, whereas other algorithms might struggle due to their adaptive learning rate mechanisms. Moreover, its stability positively affects training with smaller datasets. In such cases, the dynamic learning rates of other algorithms might cause erratic behaviour, whereas it maintains a more consistent and dependable convergence. This aligns with our need to enhance the effectiveness and reliability of our training of the classifier models. The mathematical formulae describing RMSProp can be referred from Section \ref{app:rmsprop}.

\subsubsection{Hyperparameters}
\begin{table}[ht]
\centering
\begin{tabularx}{\textwidth}{|X|c|}
\hline
Hyperparameter & Value \\
\hline
Alpha (for weighted loss) & 0.9 \\
Beta (for weighted loss) & 0.1 \\
Epochs & 150 \\
Learning Rate & 3e-5 \\
Dropout Rate & 0.4 \\
Batch Size & 128\\
\hline
\end{tabularx}
\caption{Hyperparameters for the AES model}
\label{table:hypaes}
\end{table}
The hyperparameters used for the AES model training can be glanced at from Table \ref{table:hypaes}.

\section{Competing Models}
Evaluating our proposed AES model's performance involves comparing it against several established models in the field. We consider the following three prominent competing models, each with distinct methodologies and reported achievements:

\begin{enumerate}
    \item \textbf{DeLAES Model:} DeLAES, developed by Tashu \textit{et al.} \cite{tashu2022deep}, is a novel approach that employs an n-gram-inspired CNN-Bi-GRU architecture. This model has achieved remarkable success, exhibiting a state-of-the-art mean QWK score of 0.903 across various benchmark datasets. Combining convolutional neural networks (CNNs) and bidirectional gated recurrent units (Bi-GRUs), their innovative strategy to capture local and contextual information in essays for enhanced grading accuracy along the granularities of bi-grams, tri-grams and 4-grams.
    
    \item \textbf{Tran-BERT-MS-ML-R Model:} Wang \textit{et al.} \cite{wang_2022_on} proposed the Tran-BERT-MS-ML-R model, which leverages the power of BERT-based transformer architectures. This model takes a multi-faceted approach by incorporating transfer learning techniques and specialised methodologies to improve the performance of BERT models. By addressing specific challenges associated with essay scoring, Wang \textit{et al.} achieved notable advancements in scoring accuracy.
    
    \item \textbf{$R^2$ BERT Model:} $R^2$ BERT, introduced by Yang \textit{et al.} \cite{yang-etal-2020-enhancing}, has been the state-of-the-art for transformer-based models for AES for a long time. The name $R^2$ BERT stems from its utilisation of both Regression and Ranking techniques for score prediction. This unique combination enhances the model's ability to capture intricate nuances in essays, improving grading performance. Yang \textit{et al.} demonstrated the model's efficacy through comprehensive experimentation and comparisons with existing approaches.
\end{enumerate}

These three competing models were selected based on their influential contributions to the AES domain and their notable achievements in enhancing the accuracy of essay scoring. DeLAES offers a very high state-of-the-art score for us to benchmark against. In contrast, the two BERT-based models allow us to contrast our implementation's efficacy and improvements made in transformer-based models.

\chapter{Results and Discussion}
\label{sec:results}

This chapter thoroughly discusses our study's multifaceted journey to enhance transformer-based models for Automated Essay Scoring (AES). Through a process characterised by rigorous exploration, meticulous experimentation, and thorough comparative analysis, we aimed to make meaningful contributions to the advancement of AES methodologies. Reiterating the core inquiry of this study, we have addressed the pivotal question: How can we enrich transformer-based models for AES)through the effective infusion of contextual information? We first established a standardised baseline for our experimental framework. Subsequently, we sought to explore contextual information, showing a clear scope for our exploration and probing how contextual cues could be utilised to enhance model performance.

We began by investigating the impact of truncation, a critical consideration for models utilising BERT-based architectures with token limitations. We enhanced the model's understanding by preserving the direct context of essays through non-truncation. Following this, we further delved into providing a relativistic ordering context via margin ranking, as detailed in Section \ref{sec:margin_ranking}. This entailed furnishing the model with a sense of the batch's essay score ordering to enhance its evaluation capabilities. Further, we examined the augmentation of essays with their respective prompts. This approach improved the model's grasp of essay quality and its comprehension of nuanced grading patterns within each essay set. We further explored context granularity through the levels of Elementary Discourse Unit (EDU) and Argument Component (AC). Inspired by Wang \textit{et al.}'s work \cite{wang-etal-2018-toward}, we devised models for the upstream tasks of EDU and AC span labelling. These labelled essays formed the input for our transformer-based architecture, where we observed performance improvements.

We also investigated the combined influence of prompts and AC span markers and the effects of integrating feature-based contexts directly into the embeddings, which involved contrasting different context augmentation methods and benchmarking their effectiveness against each other and the competing models. The subsequent sections will delve into the intricacies of this path, offering a detailed and exhaustive discussion of our findings and insights.

\section{Establishing Baselines}

Here, we analyse the results of our experiment aimed at establishing baseline performance using transformer-based models without additional contextual information, a fundamental reference for our subsequent investigations into context-infused models. 

\begin{table}[hbtp!]
\centering
\resizebox{\linewidth}{!}{
\begin{tabular}{|c|c|c|c|c|c|c|c|c|c|}
\hline
\textbf{Baseline Models} & \textbf{Essay Set 1} & \textbf{Essay Set 2} & \textbf{Essay Set 3} & \textbf{Essay Set 4} & \textbf{Essay Set 5} & \textbf{Essay Set 6} & \textbf{Essay Set 7} & \textbf{Essay Set 8} & \textbf{Mean QWK}\\
\hline
RoBERTa-512 Tokens & 0.8033 & 0.7117 & 0.7193 & 0.7717 & 0.7441 & 0.8264 & 0.8471 & 0.7393 & 0.7704\\
DeBERTaV3-512 Tokens & 0.6662 & 0.6328 & 0.7356 & 0.7042 & 0.8014 & 0.6225 & 0.7799 & 0.7000 & 0.7053\\
DeBERTaV3-No Truncation & 0.7660 & 0.7002 & 0.7582 & 0.7302 & 0.7965 & 0.7161 & 0.8010 & 0.7704 & 0.7548\\
Longformer-No Truncation & 0.8150 & 0.7234 & 0.7315 & 0.8384 & 0.7570 & 0.8306 & 0.8604 & 0.7123 & 0.7836\\
\hline
\end{tabular}
}
\caption{Quadratic Kappa Scores for Baseline Models}
\label{table:baseline-kappa}
\end{table}

To comprehend the inherent performance of a transformer-based model, we established a baseline using the RoBERTa architecture and a conventional classifier head. This enables us to gauge the model’s performance without any contextual augmentation to compare the baseline directly with a foundation for subsequent evaluations instead of comparing it directly with existing models in the literature. Owing to RoBERTa’s improved training techniques, it outperforms BERT, justifying our use of the former instead of the latter, as is prevalent in pre-existing research. However, we retain a standard architecture for the classifier head.
Turning to our results in Table \ref{table:baseline-kappa}, we investigate how truncation, a common strategy to handle long essays, affects model performance. We argue that RoBERTa's performance is negatively impacted by truncation limitations. This is substantiated by comparing two variations of the same architecture \textemdash \hspace{1pt} DeBERTaV3 \textemdash \hspace{1pt} with one that uses truncation and the other that does not. The non-truncated DeBERTaV3 model outperforms its truncated counterpart by approximately 7.02\%. This observation underscores the importance of preserving the entirety of an essay's context, especially considering the diverse and complex nature of essays and their varying lengths.
Having proven the importance of avoiding truncation, we examined the Longformer architecture's effect. While Longformer offers a performance improvement over RoBERTa, it was less significant than the difference between the DeBERTaV3 models. 

\section{EDU Span Classification}
As discussed before, we enhanced the transformer’s grasp of individual text segments and their relationships by leveraging the EDUs. To get EDU labels, we implemented an EDU span classification model, discussed here. Our implementations were based on Wang \textit{et al.}’s work \cite{wang-etal-2018-toward}, and we adapted their architecture to use RoBERTa embeddings instead of Glove and ElMo embeddings.  
\begin{table}[hbtp!]
\centering
\begin{tabular}{|c|c|c|c|}
\hline
\textbf{Word Embeddings} & \textbf{Precision} & \textbf{Recall} & \textbf{F1-score} \\
\hline
BiLSTM CRF with Glove and ElMo embeddings & 0.924 & 0.956 & 0.940 \\
RoBERTa without fine-tuning & 0.926 & 0.938 & 0.932 \\
Roberta with fine-tuning & 0.931 & 0.963 & 0.947 \\
\hline
\end{tabular}
\caption{Performance Metrics for EDU span labelling}
\label{table:edu}
\end{table}

We showcase our results in Table \ref{table:edu}. Using finetuning, our adapted model slightly surpasses the original (Glove and ElMO embeddings) regarding the F1-score. The well-balanced precision and recall of our top-performing EDU span prediction model underscore its proficiency in precisely detecting EDU spans. This precision is vital for downstream tasks, where these spans form the foundation for further analysis and feature extraction. The model's consistent balance hints at its potential to substantially enhance transformer-based models for Automated Essay Scoring by integrating contextual information.

\section{AC Span Classification}
This section explores the AC Span Classification task using the same model as in EDU span classification. This task aims to identify individual ACs in essays. These AC spans provide extra context for the transformer-based models in the downstream task of AES. We detail the performance of our AC Span classification model below in Table \ref{table:AC}.

\begin{table}[hbtp!]
\centering
\begin{tabular}{|c|c|c|c|}
\hline
\textbf{Word Embeddings} & \textbf{Precision} & \textbf{Recall} & \textbf{F1-score} \\
\hline
RoBERTa with Fine-Tuning & 0.910 & 0.771 & 0.835 \\
\hline
\end{tabular}
\caption{Performance Metrics for AC Span Labelling}
\label{table:AC}
\end{table}

The table provides insights into our AC span prediction task's performance. Notably, there's high precision but relatively lower recall. This suggests the model's cautious labelling approach; when it does label an AC span, it's highly accurate.  Recognising that this frugal labelling strategy could introduce noise in the AES task downstream is essential. While the model's high precision assures reliable AC spans, the lower recall implies potential omissions. Striking a better balance between precision and recall would be beneficial, ensuring accurate AC span identification and comprehensive coverage for meaningful contextual information in the AES task.

\section{Context Augmentation}
Now that we have established our baselines and implemented the models needed to provide downstream context, we will explore how these extra contextual data help with AES. The outcomes of these comprehensive experiments are documented in Table \ref{table:added_context_kappa}, which presents the Quadratic Kappa (QWK) scores attained by Longformer-based models enhanced by added context.

\begin{table}[hbtp!]
\centering
\resizebox{\linewidth}{!}{
\begin{tabular}{|c|c|c|c|c|c|c|c|c|c|}
\hline
\textbf{Models with added context} & \textbf{Essay Set 1} & \textbf{Essay Set 2} & \textbf{Essay Set 3} & \textbf{Essay Set 4} & \textbf{Essay Set 5} & \textbf{Essay Set 6} & \textbf{Essay Set 7} & \textbf{Essay Set 8} & \textbf{Mean QWK}\\
\hline
Longformer & 0.8150 & 0.7234 & 0.7315 & 0.8384 & 0.7570 & 0.8306 & 0.8604 & 0.7123 & 0.7836\\
Longformer+MR & 0.8236 & 0.7491 & 0.7311 & 0.8195 & 0.7793 & 0.8201 & 0.8593 & 0.7227 & 0.7881\\
Longformer+MR+Prompt & 0.8322 & 0.7422 & 0.7550 & 0.8521 & 0.7482 & 0.8202 & 0.8811 & 0.8319 & 0.8079\\
Longformer+MR+EDU & 0.8128 & 0.7564 & 0.7288 & 0.8175 & 0.7642 & 0.8037 & 0.8443 & 0.7968 & 0.7906\\
Longformer+MR+AC & 0.7999 & 0.7591 & 0.7825 & 0.8188 & 0.7486 & 0.8300 & 0.8526 & 0.8026 & 0.7993\\
Longformer+MR+AC+Prompt & 0.8313 & 0.7748 & 0.7704 & 0.8344 & 0.7834 & 0.8489 & 0.8742 & 0.8241 & 0.8177\\
Longformer+MR+AC+Prompt+Features  & 0.8435 & 0.7524 & 0.7670 & 0.8755 & 0.7765 & 0.8501 & 0.8701 & 0.8358 & 0.8213\\
\hline
\end{tabular}
}
\caption{Quadratic Kappa Scores for Longformer-based models with added context}
\label{table:added_context_kappa}
\end{table}

The insights derived from the results presented in Table \ref{table:added_context_kappa} unveil several significant observations regarding the influence of context augmentation on model performance across diverse essay sets:

\textbf{Model Without Additional Context:}
The baseline Longformer model shown in the table as $Longformer$, operating without any supplemental context, demonstrated comparatively diminished performance, as evidenced by its lower Quadratic Kappa scores across various essay sets.

\textbf{Margin Ranking (MR) Loss Only:}
The inclusion of Margin Ranking (MR) loss shown by $Longformer+MR$, designed to establish the relative ordering of essays, yielded a moderate performance enhancement, indicating its efficacy in capturing some nuances of essay quality.

\textbf{Prompt as Context:}
Introducing prompts as context yielded discernible performance improvement, highlighting the importance of prompt-specific insights in understanding essay characteristics and scoring variability. It is shown in the table as $Longformer+MR+Prompt$.

\textbf{Elementary Discourse Unit (EDU) as Context:}
The integration of Elementary Discourse Units (EDUs) as context contributed to a slight uptick in performance, albeit without surpassing the effectiveness of prompt context, represented by $Longformer+MR+EDU$.

\textbf{Argument Component (AC) as Context:}
ACs introduced a more substantial performance improvement than EDUs, underlining their capacity to capture critical elements of argumentation and essay structure, represented by $Longformer+MR+AC$ in the table.

\textbf{Combining ACs and Prompts:}
The fusion of Argument Components and prompts within the model architecture represented by $Longformer+MR+AC+Prompt$ further amplified performance gains, indicating their synergistic role in enhancing the model's grasp of argumentative structure and contextual nuances.

\textbf{Incorporating Feature Context:}
The final augmentation step involved integrating essay-specific feature counts, such as token counts and counts of predicted ACs and EDUs. This holistic approach yielded the best-performing model, $Longformer+MR+AC+Prompt+Features$, achieving the highest mean QWK score.

To conclude, the rigorous experimentation with context augmentation strategies has unveiled the effectiveness of these techniques in enhancing the transformer-based AES model. The amalgamation of Argument Components, prompts, and feature context yielded a model that excels across diverse essay sets, establishing a robust alignment with human graders and contributing to the advancement of AES.

\section{Comparing Models}

To assess the performance of our proposed Automated Essay Scoring (AES) model, we compare it against several established models in the field. The results of these comparisons are presented in Table \ref{table:comparing_results}, showcasing the Quadratic Kappa (QWK) scores obtained by competing models across various essay sets.

\begin{table}[hbtp!]
\centering
\resizebox{\linewidth}{!}{
\begin{tabular}{|c|c|c|c|c|c|c|c|c|c|}
\hline
\textbf{Models with added context} & \textbf{Essay Set 1} & \textbf{Essay Set 2} & \textbf{Essay Set 3} & \textbf{Essay Set 4} & \textbf{Essay Set 5} & \textbf{Essay Set 6} & \textbf{Essay Set 7} & \textbf{Essay Set 8} & \textbf{Mean QWK}\\
\hline
Tran-BERT-MS-ML-R \cite{wang_2022_on} & 0.834 & 0.716 & 0.714 & 0.812 & 0.813 & 0.836 & 0.839 & 0.766 & 0.791\\
$R^2$ BERT \cite{yang-etal-2020-enhancing} & 0.817 & 0.719 & 0.698 & 0.845 & 0.841 & 0.847 & 0.839 & 0.744 & 0.794\\
$Longformer+MR+AC+Prompt+Features$ & 0.8435 & 0.7524 & 0.7670 & 0.8755 & 0.7765 & 0.8501 & 0.8701 & 0.8358 & 0.8213\\
\hline
DeLAES \cite{tashu2022deep} & 0.927 & 0.932 & 0.884 & 0.870 & 0.925 & 0.923 & 0.887 & 0.873 & 0.903\\
$Longformer+MR+AC+Prompt+Features (with prompt-specific training)$ & 0.8981 & 0.8097 & 0.7907 & 0.8843 & 0.8959 & 0.8911 & 0.8878 & 0.9003 & 0.8697\\
\hline
\end{tabular}
}
\caption{Comparison of Quadratic Kappa Scores between competing models}
\label{table:comparing_results}
\end{table}

From the results presented in Table \ref{table:comparing_results}, a comprehensive analysis of model performance and comparisons can be made:

\begin{enumerate}
    \item \textbf{Transformer-Based BERT Models:}
Our best-performing model, $Longformer+MR+AC+Prompt+Features$, demonstrates significant improvements over two state-of-the-art transformer-based models, Tran-BERT-MS-ML-R and $R^2$ BERT, across various essay sets. These models leverage the power of transformer architectures to enhance AES, incorporating transfer learning techniques and innovative strategies to improve grading accuracy. Our model showcases its ability to outperform these established models, indicating the effectiveness of our proposed context augmentation strategies.

\item \textbf{DeLAES Model:}
While our best-performing model showcases remarkable progress, it is important to acknowledge the exceptional performance of the DeLAES model, which outperforms our model by approximately 10\%. DeLAES introduces a unique n-gram-inspired CNN-Bi-GRU architecture to capture local and contextual information, leading to its state-of-the-art mean QWK score of 0.903 across benchmark datasets.
    \begin{enumerate}
        \item \textit{Prompt-Specific Training:}
When we use prompt-specific training with our model, the results are represented by $Longformer+MR+AC+Prompt+Features (with prompt-specific training)$. It achieves competitive results in comparison to the DeLAES model. This variant of our best-performing model was trained and tested on individual essay sets to establish a fair comparison with DeLAES. The achieved QWK scores indicate that our model, with its innovative context augmentation strategies, can approach the high standards DeLAES sets in specific contexts. 
    \end{enumerate}

\end{enumerate}
The comparison of our proposed AES model against established competing models demonstrates the effectiveness of our approach in enhancing transformer-based grading accuracy, showcasing competitive performance and even outperforming certain state-of-the-art models within specific experimental setups.

\chapter{Conclusions}
\label{sec:concl}

Through the objective analysis of the observed results, we see a significant increase in the performance capabilities of transformer-based models in AES. Every scope of context augmentation implemented was reflected in increased QWK scores. Objectively,
as inferred from Table \ref{table:comparing_results}, we outperformed the previously established state-of-the-art (SOTA) for transformer-based models. Contrasting it with DeLAES, however, we still see that it beat our model by 3.83\% (approx.), even though our implementation got a higher score in some essay sets. Ostensibly, it may not be seen as a complete success, but here we re-iterate what the primary goal of our research was \textemdash \hspace{1pt}can the performance of transformer-based models be enhanced by context augmentation? The answer is yes. 
As for DeLAES outperforming our model, several reasons could be substantiated for it. Of the more direct ones \textemdash \hspace{1pt}the low recall of our AC span labelling model, using only the pre-trained weights of our transformer model, or simply the lack of a better classification head architecturally. 

Also, it should be noted that the progress made in this investigation was orthogonal to the work of Tashu \textit{et al.}. The methods of context augmentation explored can be added to any model architecture during training, whether transformer-based or not. Since the motivation would remain the same, these context augmentation techniques are also expected to increase the performance levels of non-transformer-based models.

There are rewarding avenues of future work that were not explored in this work, including:
\begin{itemize}
    \item \textbf{Argument analysis-based methods:} Ideally, we want to develop an AC span classifier with a much higher F1 score. Then, we could further explore classifying the type of AC, viz., Premise, Claim, Major Claim, and use separate special tokens for them. We could also identify the linkage between the ACs and their types and use a linearised argument graph as the input. 

    \item \textbf{Graph Convolutional Networks (GCNs):} Inspired by the work of Ruiz \textit{et al.} \cite{ruizdolz_2022_automatic}, we may even want to use GCNs to exploit the argument graph structure for further context. 

    \item \textbf{Finetuning:} Finetuning the transformer-based models is also challenging, as explained in the paper by Wang \textit{et al.} \cite{wang_2022_on}. We can explore back-translation and relativistic scoring to augment data for further training instances to stabilise fine-tuning as explored by Jong \textit{et al.} \cite{jong2022improving}.
\end{itemize}

An objective study of these would only positively benefit the literature available in AES. It is important to note that the end-to-end argument graph mining models need to achieve better SOTA results \cite{saha-etal-2022-edu} to avoid injecting noise into the transformer models instead of enriching them. 
\bibliographystyle{plain}
\bibliography{mybibfile}

\appendix

\chapter{Mathematical Formulae}
In this chapter, we discuss the mathematical formulae of different concepts discussed in the dissertation for quick reference as needed.

\section{Precision}
\label{app:pre}
Precision can be defined mathematically as: 
\[
\text{Precision} = \frac{\text{True Positives}}{\text{True Positives} + \text{False Positives}}
\]

\section{Recall}
\label{app:recall}
Recall can be defined mathematically as: 
\[
\text{Recall} = \frac{\text{True Positives}}{\text{True Positives} + \text{False Negatives}}
\]

\section{F1 Score}
\label{app:f1}
The formula for the F1 score is as follows:
\[
\text{F1 Score} = 2 \times \frac{\text{Precision} \times \text{Recall}}{\text{Precision} + \text{Recall}}
\]

\section{Negative Log Likelihood (NLL)}
\label{app:nll}
The negative log-likelihood (NLL) formula is given by:
\[
NLL(\theta) = -\sum_{i=1}^{n} \log P(x_i | \theta)
\]
where $NLL(\theta)$ represents the negative log-likelihood, $n$ is the number of data points, $x_i$ is the $i$th data point, and $P(x_i | \theta)$ is the probability of observing $x_i$ given the parameter vector $\theta$.

\section{L2 Regularisation}
\label{app:l2}
 The formula for L2 regularisation (Ridge regularisation) is given by:
\[
L2 = \lambda \cdot \|\mathbf{w}\|_2^2
\]
where $L2$ represents the L2 regularisation term, $\lambda$ is the regularisation parameter, $\mathbf{w}$ is the vector of model parameters (weights), and $\|\mathbf{w}\|_2^2$ is the squared L2 norm (Euclidean norm) of the parameter vector $\mathbf{w}$.

\section{Z-Score Normalisation}
\label{app:norm}
For a feature, Z-Score Normalisation can be mathematically shown as follows:
\[
z = \frac{x - \mu}{\sigma}
\]
where $z$ is the normalised value (Z-score) of the data point $x$, $\mu$ is the mean of the feature (score in our case) across the dataset, and $\sigma$ is the standard deviation of the feature across the dataset.

\section{MSE Loss}
\label{app:mse}
The Mean Squared Error (MSE) loss formula is given by:
\[
MSE = \frac{1}{n} \sum_{i=1}^{n} (y_i - \hat{y}_i)^2
\]
where $MSE$ represents the Mean Squared Error loss, $n$ is the number of data points, $y_i$ is the actual value of the target for the $i$th data point, and $\hat{y}_i$ is the predicted value of the target for the $i$th data point.

\section{Adam}
\label{app:adam}
Reiterating what was discussed in Section \ref{exdes:adam}, Adam keeps track of the moving averages for both the gradient and the squared gradient (referred to as the second moment). It adjusts the learning rate for each parameter by considering the magnitude of both the current gradient and the historical gradients.
The following formulae explain this mathematically:
\[
m_t = \beta_1 \cdot m_{t-1} + (1 - \beta_1) \cdot g_t, \quad
v_t = \beta_2 \cdot v_{t-1} + (1 - \beta_2) \cdot g_t^2
\]
\[
\hat{m}_t = \frac{m_t}{1 - \beta_1^t}, \quad
\hat{v}_t = \frac{v_t}{1 - \beta_2^t}
\]
\[
\theta_{t+1} = \theta_t - \frac{\alpha}{\sqrt{\hat{v}_t} + \epsilon} \cdot \hat{m}_t
\]
Where \( t \) is the iteration step, \( \alpha \) is the learning rate, \( g_t \) is the gradient at iteration \( t \), \( \beta_1 \) and \( \beta_2 \) are exponential decay rates, \( m_t \) is the first moment estimate, \( v_t \) is the second moment estimate, \( \hat{m}_t \) is the bias-corrected first moment estimate, \( \hat{v}_t \) is the bias-corrected second moment estimate, and \( \epsilon \) is a small constant.

\section{RMSProp}
\label{app:rmsprop}

The RMSProp (Root Mean Square Propagation) algorithm is a widely used optimisation technique in machine learning for training neural networks. Its purpose is to address some of the limitations of the standard gradient descent algorithm by adapting the learning rates for different parameters.

The main idea is to modify the learning rates based on the historical magnitudes of gradients. It maintains an exponentially decaying average of squared gradients for each parameter. This moving average helps take into account the fluctuation in gradients and provides a more balanced adjustment of learning rates.

The algorithm divides the current gradient by the square root of the moving average of squared gradients. This scaling factor normalises the updates, ensuring that larger gradients have a lesser impact on the update while smaller gradients exert a relatively more significant influence. This adaptive scaling assists the algorithm in navigating the optimisation landscape more effectively, leading to faster convergence towards the optimal solution.

By dynamically adjusting the learning rates in this manner, RMSProp enhances the stability and convergence speed of the optimisation process, making it particularly valuable in situations involving sparse and noisy gradients.

It can be mathematically described as follows:
\begin{align*}
E[g^2]_t &= \beta \cdot E[g^2]_{t-1} + (1 - \beta) \cdot (g_t)^2 \\
\theta_{t+1} &= \theta_t - \frac{\eta}{\sqrt{E[g^2]_t + \epsilon}} \cdot g_t
\end{align*}

Where:\\
\resizebox{0.85\linewidth}{!}{
\begin{minipage}{\linewidth}
\[
\begin{aligned}
E[g^2]_t &\text{ is the exponential moving average of squared gradients at time step } t. \\
g_t &\text{ is the cost function gradient with respect to the parameters at time step } t. \\
\theta_t &\text{ represents the parameters at time step } t. \\
\eta &\text{ is the learning rate.} \\
\beta &\text{ is the decay rate for the moving average of squared gradients (usually set to a value like 0.9).} \\
\epsilon &\text{ is a small constant added for numerical stability.}
\end{aligned}
\]
\end{minipage}
}

\chapter{Prompts in ASAP-AES Dataset}
\label{app:prompts}
This chapter gives a detailed illustration of the prompts for the different essay sets in the ASAP-AES Dataset, as follows:

\begin{itemize}
    \item \textbf{Prompt 1:} More and more people use computers, but not everyone agrees that this benefits society. Those who support advances in technology believe that computers have a positive effect on people. They teach hand-eye coordination, give people the ability to learn about faraway places and people, and even allow people to talk online with other people. Others have different ideas. Some experts are concerned that people are spending too much time on their computers and less time exercising, enjoying nature, and interacting with family and friends. 

    Write a letter to your local newspaper in which you state your opinion on the effects computers have on people. Persuade the readers to agree with you.

    \item \textbf{Prompt 2:} Censorship in the Libraries
"All of us can think of a book that we hope none of our children or any other children have taken off the shelf. But if I have the right to remove that book from the shelf -- that work I abhor -- then you also have exactly the same right and so does everyone else. And then we have no books left on the shelf for any of us." --Katherine Paterson, Author
Write a persuasive essay to a newspaper reflecting your vies on censorship in libraries. Do you believe that certain materials, such as books, music, movies, magazines, etc., should be removed from the shelves if they are found offensive? Support your position with convincing arguments from your own experience, observations, and/or reading.

\item \textbf{Prompt 3:}
Write a response that explains how the features of the setting affect the cyclist. In your response, include examples from the essay that support your conclusion.
\begin{itemize}
    \item \textit{Source Essay:} 
    ROUGH ROAD AHEAD: Do Not Exceed Posted Speed Limit\\
    by Joe Kurmaskie\\
    FORGET THAT OLD SAYING ABOUT NEVER taking candy from strangers. No, a better piece of advice for the solo cyclist would be, “Never accept travel advice from a collection of old-timers who haven’t left the confines of their porches since Carter was in office.” It’s not that a group of old guys doesn’t know the terrain. With age comes wisdom and all that, but the world is a fluid place. Things change. 
    At a reservoir campground outside of Lodi, California, I enjoyed the serenity of an early-summer evening and some lively conversation with these old codgers. What I shouldn’t have done was let them have a peek at my map. Like a foolish youth, the next morning I followed their advice and launched out at first light along a “shortcut” that was to slice away hours from my ride to Yosemite National Park.
    They’d sounded so sure of themselves when pointing out landmarks and spouting off towns I would come to along this breezy jaunt. Things began well enough. I rode into the morning with strong legs and a smile on my face. About forty miles into the pedal, I arrived at the first “town.” This place might have been a thriving little spot at one time—say, before the last world war—but on that morning it fit the traditional definition of a ghost town. I chuckled, checked my water supply, and moved on. The sun was beginning to beat down, but I barely noticed it. The cool pines and rushing rivers of Yosemite had my name written all over them. 
    Twenty miles up the road, I came to a fork of sorts. One ramshackle shed, several rusty pumps, and a corral that couldn’t hold in the lamest mule greeted me. This sight was troubling. I had been hitting my water bottles pretty regularly, and I was traveling through the high deserts of California in June.
    I got down on my hands and knees, working the handle of the rusted water pump with all my strength. A tarlike substance oozed out, followed by brackish water feeling somewhere in the neighborhood of two hundred degrees. I pumped that handle for several minutes, but the water wouldn’t cool down. It didn’t matter. When I tried a drop or two, it had the flavor of battery acid.
    The old guys had sworn the next town was only eighteen miles down the road. I could make that! I would conserve my water and go inward for an hour or so—a test of my inner spirit. 
    Not two miles into this next section of the ride, I noticed the terrain changing. Flat road was replaced by short, rolling hills. After I had crested the first few of these, a large highway sign jumped out at me. It read: ROUGH ROAD AHEAD: DO NOT EXCEED POSTED SPEED LIMIT.
    The speed limit was 55 mph. I was doing a water-depleting 12 mph. Sometimes life can feel so cruel. 
    I toiled on. At some point, tumbleweeds crossed my path and a ridiculously large snake—it really did look like a diamondback—blocked the majority of the pavement in front of me. I eased past, trying to keep my balance in my dehydrated state.
    The water bottles contained only a few tantalizing sips. Wide rings of dried sweat circled my shirt, and the growing realization that I could drop from heatstroke on a gorgeous day in June simply because I listened to some gentlemen who hadn’t been off their porch in decades, caused me to laugh.
    It was a sad, hopeless laugh, mind you, but at least I still had the energy to feel sorry for myself. There was no one in sight, not a building, car, or structure of any kind. I began breaking the ride down into distances I could see on the horizon, telling myself that if I could make it that far, I’d be fi ne.
    Over one long, crippling hill, a building came into view. I wiped the sweat from my eyes to make sure it wasn’t a mirage, and tried not to get too excited. With what I believed was my last burst of energy, I maneuvered down the hill.
    In an ironic twist that should please all sadists reading this, the building—abandoned years earlier, by the looks of it—had been a Welch’s Grape Juice factory and bottling plant. A sandblasted picture of a young boy pouring a refreshing glass of juice into his mouth could still be seen.
    I hung my head.
    That smoky blues tune “Summertime” rattled around in the dry honeycombs of my deteriorating brain.
    I got back on the bike, but not before I gathered up a few pebbles and stuck them in my mouth. I’d read once that sucking on stones helps take your mind off thirst by allowing what spit you have left to circulate. With any luck I’d hit a bump and lodge one in my throat.
    It didn’t really matter. I was going to die and the birds would pick me clean, leaving only some expensive outdoor gear and a diary with the last entry in praise of old men, their wisdom, and their keen sense of direction. I made a mental note to change that paragraph if it looked like I was going to lose consciousness for the last time.
    Somehow, I climbed away from the abandoned factory of juices and dreams, slowly gaining elevation while losing hope. Then, as easily as rounding a bend, my troubles, thirst, and fear were all behind me.
    GARY AND WILBER’S FISH CAMP—IF YOU WANT BAIT FOR THE BIG ONES, WE’RE YOUR BEST BET!
    “And the only bet,” I remember thinking.
    As I stumbled into a rather modern bathroom and drank deeply from the sink, I had an overwhelming urge to seek out Gary and Wilber, kiss them, and buy some bait—any bait, even though I didn’t own a rod or reel.
    An old guy sitting in a chair under some shade nodded in my direction. Cool water dripped from my head as I slumped against the wall beside him.
    “Where you headed in such a hurry?”
    “Yosemite,” I whispered.
    “Know the best way to get there?”
    I watched him from the corner of my eye for a long moment. He was even older than the group I’d listened to in Lodi.
    “Yes, sir! I own a very good map.”
    And I promised myself right then that I’d always stick to it in the future.
    “Rough Road Ahead” by Joe Kurmaskie, from Metal Cowboy, copyright © 1999 Joe Kurmaskie.
\end{itemize}
\item \textbf{Prompt 4:}
Read the last paragraph of the story.

"When they come back, Saeng vowed silently to herself, in the spring, when the snows melt and the geese return and this hibiscus is budding, then I will take that test again." 

Write a response that explains why the author concludes the story with this paragraph. In your response, include details and examples from the story that support your ideas.

\begin{itemize}
    \item \textit{Source Essay:} \\
    Winter Hibiscus by Minfong Ho\\
    Saeng, a teenage girl, and her family have moved to the United States from Vietnam. As Saeng walks home after failing her driver’s test, she sees a familiar plant. Later, she goes to a florist shop to see if the plant can be purchased.
    It was like walking into another world. A hot, moist world exploding with greenery. Huge flat leaves, delicate wisps of tendrils, ferns and fronds and vines of all shades and shapes grew in seemingly random profusion.
    “Over there, in the corner, the hibiscus. Is that what you mean?” The florist pointed at a leafy potted plant by the corner. 
    There, in a shaft of the wan afternoon sunlight, was a single blood-red blossom, its five petals splayed back to reveal a long stamen tipped with yellow pollen. Saeng felt a shock of recognition so intense, it was almost visceral.1
    “Saebba,” Saeng whispered.
    A saebba hedge, tall and lush, had surrounded their garden, its lush green leaves dotted with vermilion flowers. And sometimes after a monsoon rain, a blossom or two would have blown into the well, so that when she drew the well water, she would find a red blossom floating in the bucket.
    Slowly, Saeng walked down the narrow aisle toward the hibiscus. Orchids, lanna bushes, oleanders, elephant ear begonias, and bougainvillea vines surrounded her. Plants that she had not even realized she had known but had forgotten drew her back into her childhood world.
    When she got to the hibiscus, she reached out and touched a petal gently. It felt smooth and cool, with a hint of velvet toward the center—just as she had known it would feel.
    And beside it was yet another old friend, a small shrub with waxy leaves and dainty flowers with purplish petals and white centers. “Madagascar periwinkle,” its tag announced. How strange to see it in a pot, Saeng thought. Back home it just grew wild, jutting out from the cracks in brick walls or between tiled roofs.
    And that rich, sweet scent—that was familiar, too. Saeng scanned the greenery around her and found a tall, gangly plant with exquisite little white blossoms on it.  “Dok Malik,” she said, savoring the feel of the word on her tongue, even as she silently noted the English name on its tag, “jasmine.”
    One of the blossoms had fallen off, and carefully Saeng picked it up and smelled it. She closed her eyes and breathed in, deeply. The familiar fragrance filled her lungs, and Saeng could almost feel the light strands of her grandmother’s long gray hair, freshly washed, as she combed it out with the fine-toothed buffalo-horn comb. And when the sun had dried it, Saeng would help the gnarled old fingers knot the hair into a bun, then slip a dok Malik bud into it.
    Saeng looked at the white bud in her hand now, small and fragile. Gently, she closed her palm around it and held it tight. That, at least, she could hold on to. But where was the fine-toothed comb? The hibiscus hedge? The well? Her gentle grandmother? 
    A wave of loss so deep and strong that it stung Saeng’s eyes now swept over her. A blink, a channel switch, a boat ride into the night, and it was all gone. Irretrievably, irrevocably gone.
    And in the warm moist shelter of the greenhouse, Saeng broke down and wept.
    It was already dusk when Saeng reached home. The wind was blowing harder, tearing off the last remnants of green in the chicory weeds that were growing out of the cracks in the sidewalk. As if oblivious to the cold, her mother was still out in the vegetable garden, digging up the last of the onions with a rusty trowel. She did not see Saeng until the girl had quietly knelt down next to her.
    Her smile of welcome warmed Saeng. “Ghup ma laio le? You’re back?” she said cheerfully. “Goodness, it’s past five. What took you so long? How did it go? Did you—?” Then she noticed the potted plant that Saeng was holding, its leaves quivering in the wind.
    Mrs. Panouvong uttered a small cry of surprise and delight. “Dok faeng-noi!” she said. “Where did you get it?”
    “I bought it,” Saeng answered, dreading her mother’s next question.
    “How much?”
    For answer Saeng handed her mother some coins.
    “That’s all?” Mrs. Panouvong said, appalled, “Oh, but I forgot! You and the
    Lambert boy ate Bee-Maags . . . .”
    “No, we didn’t, Mother,” Saeng said.
    “Then what else—?”
    “Nothing else. I paid over nineteen dollars for it.”
    “You what?” Her mother stared at her incredulously. “But how could you? All the seeds for this vegetable garden didn’t cost that much! You know how much we—” She paused, as she noticed the tearstains on her daughter’s cheeks and her puffy eyes.
    “What happened?” she asked, more gently.
    “I—I failed the test,” Saeng said.
    For a long moment Mrs. Panouvong said nothing. Saeng did not dare look her mother in the eye. Instead, she stared at the hibiscus plant and nervously tore off a leaf, shredding it to bits.
    Her mother reached out and brushed the fragments of green off Saeng’s hands. “It’s a beautiful plant, this dok faeng-noi,” she finally said. “I’m glad you got it.”
    “It’s—it’s not a real one,” Saeng mumbled.
    “I mean, not like the kind we had at—at—” She found that she was still too shaky to say the words at home, lest she burst into tears again. “Not like the kind we had before,” she said.
    “I know,” her mother said quietly. “I’ve seen this kind blooming along the lake. Its flowers aren’t as pretty, but it’s strong enough to make it through the cold months here, this winter hibiscus. That’s what matters.”
    She tipped the pot and deftly eased the ball of soil out, balancing the rest of the plant in her other hand. “Look how root-bound it is, poor thing,” she said. “Let’s plant it, right now.”
    She went over to the corner of the vegetable patch and started to dig a hole in the ground. The soil was cold and hard, and she had trouble thrusting the shovel into it. Wisps of her gray hair trailed out in the breeze, and her slight frown deepened the wrinkles around her eyes. There was a frail, wiry beauty to her that touched Saeng deeply.
    “Here, let me help, Mother,” she offered, getting up and taking the shovel away from her.
    Mrs. Panouvong made no resistance. “I’ll bring in the hot peppers and bitter melons, then, and start dinner. How would you like an omelet with slices of the bitter melon?”
    “I’d love it,” Saeng said.
    Left alone in the garden, Saeng dug out a hole and carefully lowered the “winter hibiscus” into it. She could hear the sounds of cooking from the kitchen now, the beating of eggs against a bowl, the sizzle of hot oil in the pan. The pungent smell of bitter melon wafted out, and Saeng’s mouth watered. It was a cultivated taste, she had discovered—none of her classmates or friends, not even Mrs. Lambert, liked it—this sharp, bitter melon that left a golden aftertaste on the tongue. But she had grown up eating it and, she admitted to herself, much preferred it to a Big Mac.
    The “winter hibiscus” was in the ground now, and Saeng tamped down the soil around it. Overhead, a flock of Canada geese flew by, their faint honks clear and—yes—familiar to Saeng now. Almost reluctantly, she realized that many of the things that she had thought of as strange before had become, through the quiet repetition of season upon season, almost familiar to her now. Like the geese. She lifted her head and watched as their distinctive V was etched against the evening sky, slowly fading into the distance.
    When they come back, Saeng vowed silently to herself, in the spring, when the snows melt and the geese return and this hibiscus is budding, then I will take that test again.
    “Winter Hibiscus” by Minfong Ho, copyright © 1993 by Minfong Ho, from Join In, Multiethnic Short Stories, by Donald R. Gallo, ed.
\end{itemize}
\item \textbf{Prompt 5:}
Describe the mood created by the author in the memoir. Support your answer with relevant and specific information from the memoir.
\begin{itemize}
    \item \textit{Source Essay:} \\
    Narciso Rodriguez
from Home: The Blueprints of Our Lives
My parents, originally from Cuba, arrived in the United States in 1956. After living for a year in a furnished one-room apartment, twenty-one-year-old Rawedia Maria and twenty-seven-year-old Narciso Rodriguez, Sr., could afford to move into a modest, three-room apartment I would soon call home.
In 1961, I was born into this simple house, situated in a two-family, blond-brick building in the Ironbound section of Newark, New Jersey. Within its walls, my young parents created our traditional Cuban home, the very heart of which was the kitchen. My parents both shared cooking duties and unwittingly passed on to me their rich culinary skills and a love of cooking that is still with me today (and for which I am eternally grateful). Passionate Cuban music (which I adore to this day) filled the air, mixing with the aromas of the kitchen. Here, the innocence of childhood, the congregation of family and friends, and endless celebrations that encompassed both, formed the backdrop to life in our warm home.
Growing up in this environment instilled in me a great sense that “family” had nothing to do with being a blood relative. Quite the contrary, our neighborhood was made up of mostly Spanish, Cuban, and Italian immigrants at a time when overt racism was the norm and segregation prevailed in the United States. In our neighborhood, despite customs elsewhere, all of these cultures came together in great solidarity and friendship. It was a close-knit community of honest, hardworking immigrants who extended a hand to people who, while not necessarily their own kind, were clearly in need.
Our landlord and his daughter, Alegria (my babysitter and first friend), lived above us, and Alegria graced our kitchen table for meals more often than not. Also at the table were Sergio and Edelmira, my surrogate grandparents who lived in the basement apartment. (I would not know my “real” grandparents, Narciso the Elder and Consuelo, until 1970 when they were allowed to leave Cuba.) My aunts Bertha and Juanita and my cousins Arnold, Maria, and Rosemary also all lived nearby and regularly joined us at our table. Countless extended family members came and went — and there was often someone staying with us temporarily until they were able to get back on their feet. My parents always kept their arms and their door open to the many people we considered family, knowing that they would do the same for us.
My mother and father had come to this country with such courage, without any knowledge of the language or the culture. They came selflessly, as many immigrants do, to give their children a better life, even though it meant leaving behind their families, friends, and careers in the country they loved. They struggled both personally and financially, braving the harsh northern winters while yearning for their native tropics and facing cultural hardships. The barriers to work were strong and high, and my parents both had to accept that they might not be able to find the kind of jobs they deserved. In Cuba, Narciso, Sr., had worked in a laboratory and Rawedia Maria had studied chemical engineering. In the United States, they had to start their lives over entirely, taking whatever work they could find. The faith that this struggle would lead them and their children to better times drove them to endure these hard times.
I will always be grateful to my parents for their love and sacrifice. I’ve often told them that what they did was a much more courageous thing than I could have ever done. I’ve often told them of my admiration for their strength and perseverance, and I’ve thanked them repeatedly. But, in reality, there is no way to express my gratitude for the spirit of generosity impressed upon me at such an early age and the demonstration of how important family and friends are. These are two lessons that my parents did not just tell me. They showed me with their lives, and these teachings have been the basis of my life.
It was in this simple house that my parents welcomed other refugees to celebrate their arrival to this country and where I celebrated my first birthdays. It was in the warmth of the kitchen in this humble house where a Cuban feast (albeit a frugal Cuban feast) always filled the air with not just scent and music but life and love. It was here where I learned the real definition of “family.” And for this, I will never forget that house or its gracious neighborhood or the many things I learned there about how to love. I will never forget how my parents turned this simple house into a home.
— Narciso Rodriguez, Fashion designer
Hometown: Newark, New Jersey
“Narciso Rodriguez” by Narciso Rodriguez, from Home: The Blueprints of Our Lives. Copyright © 2006 by John Edwards.

\end{itemize}
\item \textbf{Prompt 6:}
Based on the excerpt, describe the obstacles the builders of the Empire State Building faced in attempting to allow dirigibles to dock there. Support your answer with relevant and specific information from the excerpt.
\begin{itemize}
    \item  \textit{Source Essay:} \\
    The Mooring Mast
by Marcia Amidon Lüsted
When the Empire State Building was conceived, it was planned as the world’s tallest building, taller even than the new Chrysler Building that was being constructed at Forty-second Street and Lexington Avenue in New York. At seventy-seven stories, it was the tallest building before the Empire State began construction, and Al Smith was determined to outstrip it in height.
The architect building the Chrysler Building, however, had a trick up his sleeve. He secretly constructed a 185-foot spire inside the building, and then shocked the public and the media by hoisting it up to the top of the Chrysler Building, bringing it to a height of 1,046 feet, 46 feet taller than the originally announced height of the Empire State Building.
Al Smith realized that he was close to losing the title of world’s tallest building, and on December 11, 1929, he announced that the Empire State would now reach the height of 1,250 feet. He would add a top or a hat to the building that would be even more distinctive than any other building in the city. John Tauranac describes the plan:
[The top of the Empire State Building] would be more than ornamental, more than a spire or dome or a pyramid put there to add a desired few feet to the height of the building or to mask something as mundane as a water tank. Their top, they said, would serve a higher calling. The Empire State Building would be equipped for an age of transportation that was then only the dream of aviation pioneers.
This dream of the aviation pioneers was travel by dirigible, or zeppelin, and the Empire State Building was going to have a mooring mast at its top for docking these new airships, which would accommodate passengers on already existing transatlantic routes and new routes that were yet to come.
The Age of Dirigibles
By the 1920s, dirigibles were being hailed as the transportation of the future. Also known today as blimps, dirigibles were actually enormous steel-framed balloons, with envelopes of cotton fabric filled with hydrogen and helium to make them lighter than air. Unlike a balloon, a dirigible could be maneuvered by the use of propellers and rudders, and passengers could ride in the gondola, or enclosed compartment, under the balloon.
Dirigibles had a top speed of eighty miles per hour, and they could cruise at seventy miles per hour for thousands of miles without needing refueling. Some were as long as one thousand feet, the same length as four blocks in New York City. The one obstacle to their expanded use in New York City was the lack of a suitable landing area. Al Smith saw an opportunity for his Empire State Building: A mooring mast added to the top of the building would allow dirigibles to anchor there for several hours for refueling or service, and to let passengers off and on. Dirigibles were docked by means of an electric winch, which hauled in a line from the front of the ship and then tied it to a mast. The body of the dirigible could swing in the breeze, and yet passengers could safely get on and off the dirigible by walking down a gangplank to an open observation platform.
The architects and engineers of the Empire State Building consulted with experts, taking tours of the equipment and mooring operations at the U.S. Naval Air Station in Lakehurst, New Jersey. The navy was the leader in the research and development of dirigibles in the United States. The navy even offered its dirigible, the Los Angeles, to be used in testing the mast. The architects also met with the president of a recently formed airship transport company that planned to offer dirigible service across the Pacific Ocean.
When asked about the mooring mast, Al Smith commented:
[It’s] on the level, all right. No kidding. We’re working on the thing now. One set of engineers here in New York is trying to dope out a practical, workable arrangement and the Government people in Washington are figuring on some safe way of mooring airships to this mast.
Designing the Mast
The architects could not simply drop a mooring mast on top of the Empire State Building’s flat roof. A thousand-foot dirigible moored at the top of the building, held by a single cable tether, would add stress to the building’s frame. The stress of the dirigible’s load and the wind pressure would have to be transmitted all the way to the building’s foundation, which was nearly eleven hundred feet below. The steel frame of the Empire State Building would have to be modified and strengthened to accommodate this new situation. Over sixty thousand dollars’ worth of modifications had to be made to the building’s framework.
Rather than building a utilitarian mast without any ornamentation, the architects designed a shiny glass and chrome-nickel stainless steel tower that would be illuminated from inside, with a stepped-back design that imitated the overall shape of the building itself. The rocket-shaped mast would have four wings at its corners, of shiny aluminum, and would rise to a conical roof that would house the mooring arm. The winches and control machinery for the dirigible mooring would be housed in the base of the shaft itself, which also housed elevators and stairs to bring passengers down to the eighty-sixth floor, where baggage and ticket areas would be located.
The building would now be 102 floors, with a glassed-in observation area on the 101st floor and an open observation platform on the 102nd floor. This observation area was to double as the boarding area for dirigible passengers.
Once the architects had designed the mooring mast and made changes to the existing plans for the building’s skeleton, construction proceeded as planned. When the building had been framed to the 85th floor, the roof had to be completed before the framing for the mooring mast could take place. The mast also had a skeleton of steel and was clad in stainless steel with glass windows. Two months after the workers celebrated framing the entire building, they were back to raise an American flag again—this time at the top of the frame for the mooring mast.
The Fate of the Mast
The mooring mast of the Empire State Building was destined to never fulfill its purpose, for reasons that should have been apparent before it was ever constructed. The greatest reason was one of safety: Most dirigibles from outside of the United States used hydrogen rather than helium, and hydrogen is highly flammable. When the German dirigible Hindenburg was destroyed by fire in Lakehurst, New Jersey, on May 6, 1937, the owners of the Empire State Building realized how much worse that accident could have been if it had taken place above a densely populated area such as downtown New York.
The greatest obstacle to the successful use of the mooring mast was nature itself. The winds on top of the building were constantly shifting due to violent air currents. Even if the dirigible were tethered to the mooring mast, the back of the ship would swivel around and around the mooring mast. Dirigibles moored in open landing fields could be weighted down in the back with lead weights, but using these at the Empire State Building, where they would be dangling high above pedestrians on the street, was neither practical nor safe.
The other practical reason why dirigibles could not moor at the Empire State Building was an existing law against airships flying too low over urban areas. This law would make it illegal for a ship to ever tie up to the building or even approach the area, although two dirigibles did attempt to reach the building before the entire idea was dropped. In December 1930, the U.S. Navy dirigible Los Angeles approached the mooring mast but could not get close enough to tie up because of forceful winds. Fearing that the wind would blow the dirigible onto the sharp spires of other buildings in the area, which would puncture the dirigible’s shell, the captain could not even take his hands off the control levers. 
Two weeks later, another dirigible, the Goodyear blimp Columbia, attempted a publicity stunt where it would tie up and deliver a bundle of newspapers to the Empire State Building. Because the complete dirigible mooring equipment had never been installed, a worker atop the mooring mast would have to catch the bundle of papers on a rope dangling from the blimp. The papers were delivered in this fashion, but after this stunt the idea of using the mooring mast was shelved. In February 1931, Irving Clavan of the building’s architectural office said, “The as yet unsolved problems of mooring air ships to a fixed mast at such a height made it desirable to postpone to a later date the final installation of the landing gear.”
By the late 1930s, the idea of using the mooring mast for dirigibles and their passengers had quietly disappeared. Dirigibles, instead of becoming the transportation of the future, had given way to airplanes. The rooms in the Empire State Building that had been set aside for the ticketing and baggage of dirigible passengers were made over into the world’s highest soda fountain and tea garden for use by the sightseers who flocked to the observation decks. The highest open observation deck, intended for disembarking passengers, has never been open to the public.
“The Mooring Mast” by Marcia Amidon Lüsted, from The Empire State Building. Copyright © 2004 by Gale, a part of Cengage Learning, Inc.

\end{itemize}
\item \textbf{Prompt 7:}
Write about patience. Being patient means that you are understanding and tolerant. A patient person experience difficulties without complaining.
Do only one of the following: write a story about a time when you were patient OR write a story about a time when someone you know was patient OR write a story in your own way about patience.

\item \textbf{Prompt 8:}
We all understand the benefits of laughter. For example, someone once said, “Laughter is the shortest distance between two people.” Many other people believe that laughter is an important part of any relationship. Tell a true story in which laughter was one element or part.

\end{itemize}


\end{document}